\documentclass[prx,twocolumn,longbibliography,superscriptaddress,preprintnumbers]{revtex4-2}
\usepackage[colorlinks,bookmarks=true,citecolor=blue,linkcolor=blue,urlcolor=blue]{hyperref}
\usepackage{dcolumn,graphicx,amsfonts,amsthm,bm,color,appendix,float}
\usepackage{braket}
\usepackage[normalem]{ulem}
\usepackage[version=4]{mhchem}
\usepackage{siunitx}
\usepackage{amsmath}
\usepackage{amssymb}

\usepackage[dvipsnames]{xcolor}
\definecolor{pal0}{rgb}{0.8941, 0.102 , 0.1098}
\definecolor{pal1}{rgb}{0.2157, 0.4941, 0.7216}
\definecolor{pal2}{rgb}{0.302 , 0.6863, 0.2902}
\definecolor{pal3}{rgb}{0.5961, 0.3059, 0.6392}
\definecolor{pal4}{rgb}{1.    , 0.498 , 0.    }

\newcommand{\old}[1]{{ \color{gray}[old] #1}}
\DeclareMathOperator{\Tr}{Tr}

\newcommand{\n}[1]{\left| #1 \right|}
\newcommand{\st}[1]{\left\{#1\right\}}
\renewcommand{\v}[1]{\boldsymbol{#1}}

\begin{document}

\title{Composite Fermi Liquid at Zero Magnetic Field in Twisted MoTe$_2$}

\author{Junkai Dong}
\email{junkaidong@g.harvard.edu}
\affiliation{Department of Physics, Harvard University, Cambridge, MA 02138, USA}

\author{Jie Wang}
\email{jiewang@fas.harvard.edu}
\affiliation{Department of Physics, Harvard University, Cambridge, MA 02138, USA}
\affiliation{Center of Mathematical Sciences and Applications, Harvard University, Cambridge, MA 02138, USA}

\author{Patrick J. Ledwith}
\email{pledwith@g.harvard.edu}
\affiliation{Department of Physics, Harvard University, Cambridge, MA 02138, USA}

\author{Ashvin Vishwanath}
\email{avishwanath@g.harvard.edu}
\affiliation{Department of Physics, Harvard University, Cambridge, MA 02138, USA}

\author{Daniel E. Parker}
\email{daniel\_parker@fas.harvard.edu}
\affiliation{Department of Physics, Harvard University, Cambridge, MA 02138, USA}

\begin{abstract}

The pursuit of exotic phases of matter outside of the extreme conditions of a quantizing magnetic field is a longstanding quest of solid state physics. Recent experiments have observed spontaneous valley polarization and fractional Chern insulators (FCIs) in zero magnetic field in twisted bilayers of MoTe$_2$, at partial filling of the topological valence band ($\nu=-2/3 $ and $-3/5$).   We study the topological valence band at {\em half} filling, using exact diagonalization and density matrix renormalization group calculations. We discover a  composite Fermi liquid (CFL) phase even at zero magnetic field that covers a large portion of the phase diagram near twist angle ${\sim}3.6^\circ$. The CFL is a non-Fermi liquid phase with metallic behavior despite the absence of Landau quasiparticles. We discuss experimental implications including the competition between the CFL and a Fermi liquid, which can be tuned with a displacement field.  The topological valence band has excellent quantum geometry over a wide range of twist angles and a small bandwidth that is, remarkably, reduced by interactions. These key properties stablize the exotic zero field quantum Hall phases. Finally, we present an optical signature involving ``extinguished" optical responses that detects Chern bands with ideal quantum geometry. 
\end{abstract}

\maketitle

Strong interactions can lead to exotic phases of matter such as non-Fermi liquids. A remarkable example is the composite Fermi liquid (CFL) that occurs in a half or quarter filled lowest Landau level (LLL). The CFL is a non-Fermi liquid with an emergent Fermi sea composed of charge neutral ``composite fermions''~\cite{JainCF89,LopezFradkin,HalperinLeeRead,Son} and has anomalous responses to a wide variety of experimental probes~\cite{STMCFLEisenstein,WillettPRL,KangPRL,SAWWillett,Goldman1994MagneticFocusing,Smet1996Focusing}. The gapless CFL state has provided an elegant interpretation for various Abelian~\cite{JainCF89,LopezFradkin,HalperinLeeRead,Son} and non-Abelian gapped topological phases~\cite{ReadGreen20}. 

This work proposes an alternative route to realize CFLs. Our proposal is based on twisted 2D transition metal dichalcogenides (TMD), a family of platforms that have realized a wealth of interesting phenomena~\cite{Tang:2020aa,Xu_2022,Wang_2020,Xu:2020aa,Huang_2021,Regan:2020aa,Li_2021,Ghiotto_2021,Li_2021_2,zhao2022realization,MoTe2WSe2,Foutty_2023,foutty2023mapping,anderson2023programming,FCITMD23_xiaodong,FCITMD23_Kinfai}, and generated much theoretical interest for their topological properties~\cite{MacDonald1,wu2019topological,YaoTMD1,YaoTMD2,Tang_2021, LiangMagic,LiangTMD6,Zhang_2021, MillisTMD1,MillisTMD2,MillisTMD3,PhysRevResearch.2.033087,abouelkomsan2022multiferroicity,crepel2023chiral}. A recent experiment~\cite{FCITMD23_xiaodong} provided strong evidence for zero field fractional Chern insulators (FCIs)~\cite{Parameswaran_2013, Neupertreview_2015,zhao_review,Liu_review_2023} at fillings $\nu=-2/3$ and $-3/5$ in twisted bilayer \ce{MoTe2} (t\ce{MoTe2}). The $\nu=-2/3$ FCI was separately found by Ref.~\cite{FCITMD23_Kinfai}. These experiments were preceded by theoretical models of Chern bands in t\ce{MoTe2}~\cite{wu2019topological}, as well as numerical works that found FCIs at partial fillings in \ce{MoTe2}~\cite{Kaisun_FCI21} and in \ce{WSe2}~\cite{Valentin22_anomaloushallmetal,Cano_pressure_23}. More recently, theoretical studies combining \textit{ab-initio} lattice relaxation and exact diagonalization on t\ce{MoTe2}~\cite{FCITMD23_Dixiao,FCITMD23_LF} 
have also obtained FCIs.

FCIs were previously reported at high magnetic fields~\cite{SpantonFCI} by partially filling Hofstadter bands~\cite{kolread1993} of a substrate-induced moir\'e potential in graphene. Shortly thereafter, with the discovery of correlated phenomena~\cite{Cao:2018ab, Cao:2018aa} and spontaneous Chern insulators~\cite{Senthil_NearlyFlatBand, Sharpe605, Serlin900} in twisted bilayer graphene (TBG), FCIs in zero field were theoretically anticipated in magic-angle TBG~\cite{Grisha_TBG2, Abouelkomsan_ph22,Cecile_TBG_Flatband}. Experimental observations of FCIs in this setting soon appeared ~\cite{FCI_TBG_exp}, albeit in a small magnetic field that theory~\cite{parker2021field} found was needed to improve the bandwidth and quantum geometry. These barriers are strikingly absent in t\ce{MoTe2}, motivating us to go beyond zero field FCIs to an exotic gapless state --- the CFL.

We will focus on the gapless CFL phase, which presents challenges~\cite{GeraedtsScience,Ippoliti1,Ippoliti2,Ippoliti3} relative to the well-understood spectral and entanglement signatures present in gapped FCI phases~\cite{KaiSunNearlyFlat,Sheng:2011tr,NeupertFQHZeroField,WangFQHBoson,TangHighTemperature, PhysRevX.1.021014}.
Combining large scale exact diagonalization (ED) with density matrix renormalization group (DMRG) numerics, we find a broad CFL phase at experimentally realistic parameters of t\ce{MoTe2}.
Furthermore, we present an explicit trial wavefunction that captures the essential features of the zero field CFL and its low energy spectrum.
Finally, we discuss experimental signatures that distinguish the CFL from Fermi liquids, enabling experimental exploration.

\begin{figure}[t]
\centering
\includegraphics[width=\linewidth]{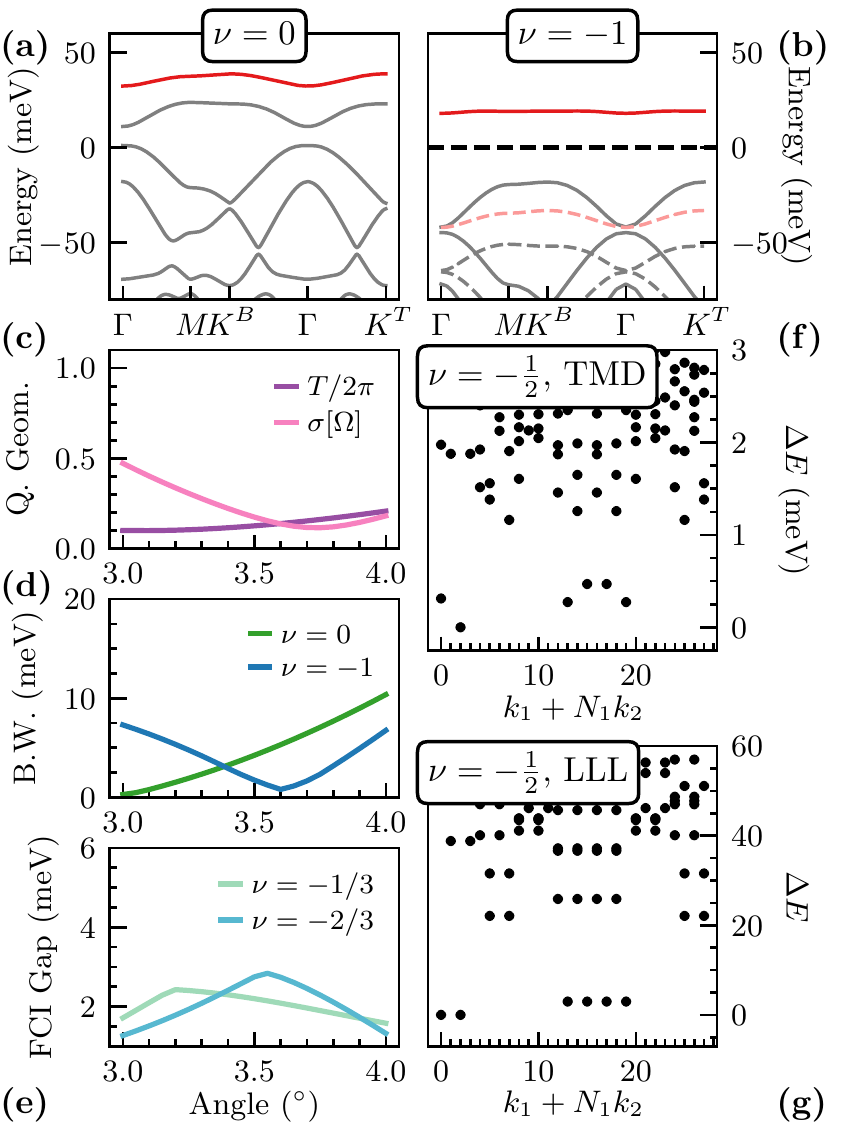}
\caption{The top valence band has favorable conditions for fractionalized topological phases. Bandstructure as seen from (a) charge neutrality and (b) from $\nu=-1$ computed from self-consistent Hartree-Fock. 
(c) Quantum geometry in terms of trace condition $T$ and Berry curvature deviation $\sigma[\Omega]$. 
(d) Bare and SCHF bandwidths and (e) the many-body gap of FCIs at $\nu=-1/3$ and $\nu=-2/3$ as a function of twist angle. The FCI gaps are obtained from ED with $N_e=8$ and $16$ respectively. 
(f) and (g): ED spectrum for 14 particles at half filling for Coulomb interaction in lowest Landau level and screened Coulomb interaction in twisted \ce{MoTe2}, respectively. 
Parameters: $(\theta, \epsilon_r, d) = (3.7^\circ, 15, \SI{300}{\angstrom})$ unless otherwise noted.}
\label{fig:bandstructure_geometry_optics}
\end{figure}

\emph{Continuum model.---}We consider a  model~\cite{wu2019topological}  for the valence bands of a twisted TMD with gate-screened~\cite{Zaletel_TBG_AQH} Coulomb interactions
\begin{equation}
\hat{H} = -\hat{h} + \frac{1}{2A}\sum_{\v{q}} V_{\v{q}} :\hat{\rho}_{\v{q}} \hat{\rho}_{-\v{q}}:,\,\,\, V_{q} = \frac{2\pi \tanh(q d)}{\epsilon_r \epsilon_0 q},
\label{eq:TMD_hamiltonian}
\end{equation}
where $\hat{\rho}_{\v q}$ is the density operator, $A$ is the sample area, normal ordering is relative to filling $\nu=0$, $d$ is gate distance, and $\epsilon_r~\approx~8-40$ is the dielectric constant~\cite{FCITMD23_Dixiao}. Due to spin-valley locking~\cite{wu2019topological}, the low energy holes of the $K$ ($K'$) valley are locked to spin up (down). The total kinetic term is $h = h_{K} + h_{K'}$ with~\cite{wu2019topological}
\begin{equation}
    h_K = \begin{bmatrix} h^b(\v{r})+V/2 & T(\bm r) \\ T^\dag(\bm r) & h^t(\v{r})-V/2\end{bmatrix},
    \label{eq:hole_bandstructure}
\end{equation}
where $h^\ell(\v{r}) = - (\v{p} - \hbar v_F \v{K}^\ell)^2/{2m^*}  + \Delta^\ell(\v{r})$ and $h_{K'}$ is determined by time-reversal.
Here the layer-diagonal terms include the quadratic monolayer TMD dispersion centered at rotated monolayer $K$-points $\v{K}^{t/b}$, shifted by the displacement field $V$, and the moir\'e potentials $\Delta^{b/t}(\bm r) = 2v\sum_{j=1,3,5}\cos\left(\bm b_j\cdot\bm r\pm\psi\right)$. The off-diagonal terms are interlayer tunnelings $T(\bm r) = \omega\left(1+e^{i\bm b_2\cdot\bm r}+e^{i\bm b_3\cdot\bm r}\right)$, where $\v{b}_j$ are the reciprocal vectors obtained by counterclockwise $(j-1)\pi/3$ rotations of $\v{b}_1 = (4\pi 3^{-1/2} \theta/ a_0, 0)$. We focus on t\ce{MoTe2}, where recent first-principles calculations~\cite{FCITMD23_Dixiao} (see also~\cite{FCITMD23_LF,wu2019topological}) found $(a_0, m^*, V,\psi,\omega) = (\SI{3.52}{\angstrom},0.6m_e, \SI{20.8}{\milli\electronvolt}, -107.7^\circ, \SI{-23.8}{\milli\electronvolt})$. We take $\theta=3.7^\circ$ throughout.

\emph{Flat Almost-Ideal Chern Band.---}Fig.~\ref{fig:bandstructure_geometry_optics}(a) shows the bandstructure for electrons $h_K$. The top moir\'e band has Chern number $C=1$, due to the skyrmionic character of the layer spinor~\cite{wu2019topological}. 

Recent experiments~\cite{FCITMD23_xiaodong,FCITMD23_Kinfai} demonstrate that the many-body ground state is ferromagnetic (valley-polarized) in at least the range $-1.2 \lesssim \nu \lesssim -0.4$. The ``parent state" for this regime is the correlated insulating state at $\nu=-1$. Fig.~\ref{fig:bandstructure_geometry_optics}(b) shows its bandstructure within self-consistent Hartree-Fock (SCHF), which is strongly renormalized by interactions. Strikingly, the renormalized $C=1$ band (red) becomes almost exactly flat, with bandwidth $\SI{1.6}{\milli\electronvolt}$ at $\theta=3.7^\circ$. This \textit{reduction}~\cite{grushinEnhancingStabilityFractional2012} in bandwidth from interaction effects is highly unusual~\cite{footnote1}.%

The many-body physics of such flat bands is determined by the Bloch wavefunctions, often through their ``quantum geometry''. Recent theories~\cite{PhysRevB.90.165139,Parameswaran_2013,Jackson:2015aa,Martin_PositionMomentumDuality,Grisha_TBG,Grisha_TBG2,kahlerband1,kahlerband2,kahlerband3,JieWang_exactlldescription,Jie_hierarchyidealband,LedwithVishwanathParker22,JW_origin_22,junkaidonghighC22} emphasize the role of K\"ahler geometry in FCI stability. We say that a band has ``ideal quantum geometry" if the trace inequality $T = \int d^2 \v k \, (\Tr g_{\text{FS}}(\v k) - \Omega(\v k)) \geq 0$ is saturated~\cite{Grisha_TBG, Grisha_TBG2, JieWang_exactlldescription,Emil_constantBerry}; here $g_{\text{FS}}$ is the Fubini-Study metric and $\Omega$ is the Berry curvature. Ideal bands are ``vortexable'' in the sense that $\hat{z} P = P \hat{z} P$ where $P$ is the projector onto the band and $\hat{z} = \hat{x}+i\hat{y}$ ~\cite{LedwithVishwanathParker22,Eslam_highC_idealband}. Vortexability enables the direct construction of Laughlin-like FQHE trial states that are exact many-body ground states for ideal bands with short-range interactions~\cite{Eslam_highC_idealband,LedwithVishwanathParker22,TrugmanKivelson85}. Fig. \ref{fig:bandstructure_geometry_optics}(c) shows $T$, the deviation from ideality, and $\sigma[\Omega]$, the standard deviation of Berry curvature. Both are small in t\ce{MoTe2} for $3^\circ \le \theta \le 4^\circ$. The top valence band thus has the rare combination of excellent quantum geometry and negligible bandwidth that favors lattice realizations of exotic quantum Hall states at zero magnetic field.

The interacting physics of the flat band is modelled by projecting Eq.~\eqref{eq:TMD_hamiltonian} via $-\hat{h} \to \sum_{\v k} \epsilon(\v k) \hat{c}^\dag_{\v k} \hat{c}_{\v k} $ and $\hat{\rho}_{\v q} \to \overline{\rho}_{\v q} = \sum_{\v k} \hat{c}^\dag_{\v k} \langle u_{\v k} | u_{\v k + \v q} \rangle \hat{c}_{\v k + \v q} $ where $\epsilon(\v k)$ and $u_{\v k}$ are the dispersion and periodic part of Bloch wavefunction. Fig. \ref{fig:bandstructure_geometry_optics}(d) shows the bare ($\nu=0$) and renormalized ($\nu=-1$) bandwidths versus twist angle, minimized near $3^\circ$ and $3.6^\circ$, respectively. Fig. \ref{fig:bandstructure_geometry_optics}(e) confirms that FCIs are stabilized at $\nu=-1/3,-2/3$ --- in concord with previous results~\cite{Kaisun_FCI21,FCITMD23_Dixiao,FCITMD23_LF}. The mild angular dependence should make FCIs relatively robust to twist angle disorder. Notably the gap at $\nu=-2/3$ is largest where the bandwidth at $\nu=-1$ is smallest~\cite{footnote2}. We therefore expect ${\sim}3.6^\circ$ to be optimal for FQH physics at half filling.

\begin{figure}
    \centering
    \includegraphics[width=\linewidth]{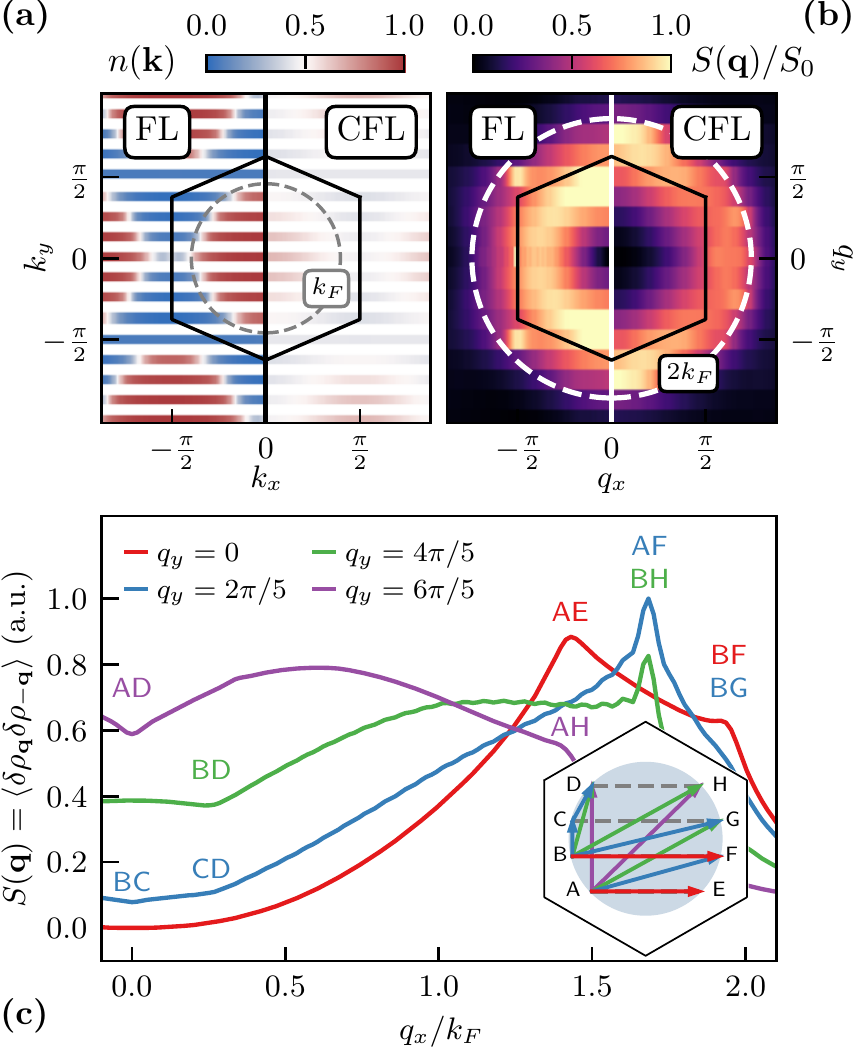}
    \caption{Numerical identification of the composite Fermi liquid (CFL) from iDMRG. (a) Occupations $n(\v{k})$ in the Brillouin zone at $L_y=8$ for the Fermi liquid (FL, left side) versus the CFL (right side).
    (b) Connected structure factor $S(\v{q})=\braket{\hat{\rho}_{\v{q}}\hat{\rho}_{-\v{q}}} - \braket{\hat{\rho}_{\v{q}}} \braket{\hat{\rho}_{-\v{q}}}$ at $L_y=8$. Characteristic features of a Fermi surface are visible for both the FL and CFL: near-vanishing weight outside $|\v{q}|\approx 2k_F$, and peaks corresponding to momentum transfers inside that radius.
    (c) Cuts of $S(\v{q})$ at constant $q_y$ for $L_y=5$ for the CFL. Each peak or inflection in $S(\v{q})$ quantitatively matches scattering events across the almost-circular composite Fermi surface (Inset). Parameters match Fig. \ref{fig:bandstructure_geometry_optics} with $\epsilon_r = 15$ ($100$) for the CFL (FL).}
    \label{fig:CFL_numerics}
\end{figure}

\emph{Composite Fermi liquid at $\nu=-1/2$.---}
We now go beyond gapped FCIs and examine the more exotic gapless CFL state~\cite{JainCF89,ReadGreen20}. We focus on $\nu=-1/2$ but our conclusions also apply to $\nu=-3/4$ (data in SM). 

{\it (i) Many body spectrum:} Fig. \ref{fig:bandstructure_geometry_optics}(f, g) compares the spectra of twisted \ce{MoTe2} and the lowest Landau level (LLL) at half filling with $14$ electrons, showing a one-to-one correspondence at low energy. The LLL spectrum uses the same geometry as t\ce{MoTe2} with Coulomb interactions. This one-to-one similarity holds at all system sizes $N_e = 8 - 14$. We thus conclude that the ground state of $\hat{H}$ at $\nu=-1/2$ is the same phase as the half-filled LLL with Coulomb interactions --- the CFL. The ground state and low-energy excitations are at precisely the momenta expected for compact composite Fermi sea (CFS) configurations~\cite{scottjiehaldane}. See SM for other system sizes, and detailed matching of degeneracies, momenta, and excitations to CFL expectations.

{\it (ii) Absence of electron Fermi surface:} A finite quasiparticle weight $Z>0$ gives the jump in electron occupations $n(\v{k})$ at the Fermi surface in a regular Fermi liquid (FL). As a non-Fermi liquid, composite fermions have vanishing $Z$, leading to the absence of Fermi surface occupation discontinuities. 

To characterize the CFL, we employ large-scale iDMRG~\cite{white1992density,schollwock2011density} calculations with the TenPy library~\cite{tenpy}. We use an infinite cylinder geometry with circumference $L_y = 5-10$, corresponding to $L_y$ evenly spaced horizontal wires through the Brillouin zone (Fig. \ref{fig:CFL_numerics}(c) inset). We take a computational basis of hybrid Wannier orbitals~\cite{marzari-vanderbilt, vanderbilt2018berry,WannierQi}, and use ``MPO compression''~\cite{MPOCompression1,MPOCompression2} to accurately capture gate-screened Coulomb interactions in the flat band. Under weak interactions ($\epsilon_r = 100$), we find the FL expected from band theory at $\nu=-1/2$, with an almost-circular Fermi surface centered at $\Gamma$ (Fig. \ref{fig:CFL_numerics}(a), left) with radius $k_F = (A_{\text{BZ}}/2\pi)^{1/2}$. The SM shows electrons, holes, and particle-hole pairs are likely gapless~\cite{pollmann2012detection}, confirming the Fermi liquid.

Under realistic interactions ($\epsilon_r=15$) with the same parameters, the ground state has quasi-uniform occupations $|n(\v{k}) - \frac{1}{2}| < 0.17$ (Fig. \ref{fig:CFL_numerics}(a), right). Because charge $Q_E = 1$ correlations are short-ranged, the state is inconsistent with an electronic Fermi liquid. However, the state has high entanglement and significant electrically-neutral correlations, consistent with the gapless density fluctuations expected from an emergent CFS. To reveal the ``hidden'' CFS, we turn to the structure factor.

{\it (iii) Scattering across the composite Fermi sea:} Fig.~\ref{fig:CFL_numerics}(b) contrasts the connected structure factor $S(\v{q})=\braket{\hat{\rho}_{\v{q}}\hat{\rho}_{-\v{q}}} - \braket{\hat{\rho}_{\v{q}}} \braket{\hat{\rho}_{-\v{q}}}$ between the FL and the CFL.  Both nearly vanish when $|\v{q}|>2k_F$, strongly implying that there is a Fermi surface in the CFL phase whose constituent fermions aren't electrons. We then match the features of $S(\v{q})$ to scattering events with different momentum transfers across the putative CFS in Fig.~\ref{fig:CFL_numerics}(c), e.g. $\hat{c}_{\v{k}=G}^\dagger \hat{c}_{\v{k}=B}$ scattering with $q_x \approx 1.94 k_F$.
The \textit{tour-de-force} work of Geraedts \textit{et al}~\cite{GeraedtsScience} showed such features are emblematic of the CFS arising from the half-filled LLL. As every feature in $S(\v{q})$ corresponds to such a scattering (quantitative matching in SM), we conclude the state has an almost-circular~\cite{footnote3} 
CFS composed of non-Landau quasiparticles. 
These two independent numerical methods establish a CFL state at $\nu=-1/2$ (see SM for $\nu=-3/4$).

\textit{Zero Field CFL Wavefunction.---} Standard theories of composite fermions apply at $B>0$, where emergent gauge flux cancels external magnetic flux. These cannot apply directly here at zero magnetic field. We therefore construct an explicit zero-field CFL wavefunction. To start, we approximate the geometry of the top t\ce{MoTe2} band as ideal. Such bands have the general ``LLL-like" wavefunction~\cite{JieWang_exactlldescription,Grisha_TBG2},
\begin{equation}
    \psi_{l}(\v r) = \phi(\v r) \zeta_l(\v r)  = f(z) e^{-K(\v r)} \zeta_l(\v r),
    \label{eq:singleparticle_generalform}
\end{equation}
where $f(z)$ is holomorphic and $\zeta_l(\v r)$ is an orbital-space spinor where $\sum_l |\zeta_l(\v r)|^2 = 1$. Here $\phi(\v r)$ is the wavefunction of a Dirac particle in an inhomogeneous, periodic, magnetic field $B(\v r) = \nabla^2 \text{Re} K(\v r)$ with one flux per unit cell~\cite{Grisha_TBG2,Liang_DiracNonuniformB,valentin23ideal}. While $\psi$ is symmetric under ordinary translations, $\phi(\v r)$ and $\zeta_l(\v r)$ are symmetric under \textit{magnetic} translations, with opposite magnetic twists~\cite{JieWang_NodalStructure},  giving a gauge redundancy $\phi(\v r) \to e^{+i \lambda(\v r)} \phi(\v r), \zeta_l(\v r) \to e^{-i \lambda(\v r)} \zeta_l(\v r)$. The form Eq.\eqref{eq:singleparticle_generalform} implies that \emph{all} many-body wavefunctions within the band of interest have the form $\Psi = \Psi_\phi \prod_i \zeta_{l_i}(\v r_i)$, where $\Psi_\phi$ is a wavefunction of flux-feeling particles; in the SM we interpret this fractionalization in terms of a new type of Chern band parton theory~\cite{Jain1989incompressible,wen1991nonabelian}; see also~\cite{luSymmetryprotectedFractionalChern2012a,mcgreevyWaveFunctionsFractional2012,murthyCompositeFermionsFractionally2011,murthyHamiltonianTheoryFractionally2012,barkeshliContinuousTransitionsComposite2012,barkeshliContinuousTransitionFractional2014}. For example, we may use Read \& Rezayi's LLL ansatz for the CFL~\cite{rezayiFermiliquidlikeStateHalffilled1994} to obtain:
\begin{equation}
    \Psi(\{ \v r_i \}) = \mathcal{P} \det_{ij} \psi^{\text{CF}}_{\v k_i}(\v r_j)\prod_{i<j} (z_i - z_j)^2 \prod_i e^{-K(\v r_i)} \zeta_{l_i}(\v r_i).
    \label{eq:CFL_chern_wavefunction}
\end{equation}
Here $\mathcal{P} = \prod_i P_i$ is the many-body projector to the top band, and $\psi^{\text{CF}}_{\v{k}_i}$ fill a Fermi sea~\cite{footnote4}.

\begin{figure}
    \centering    \includegraphics{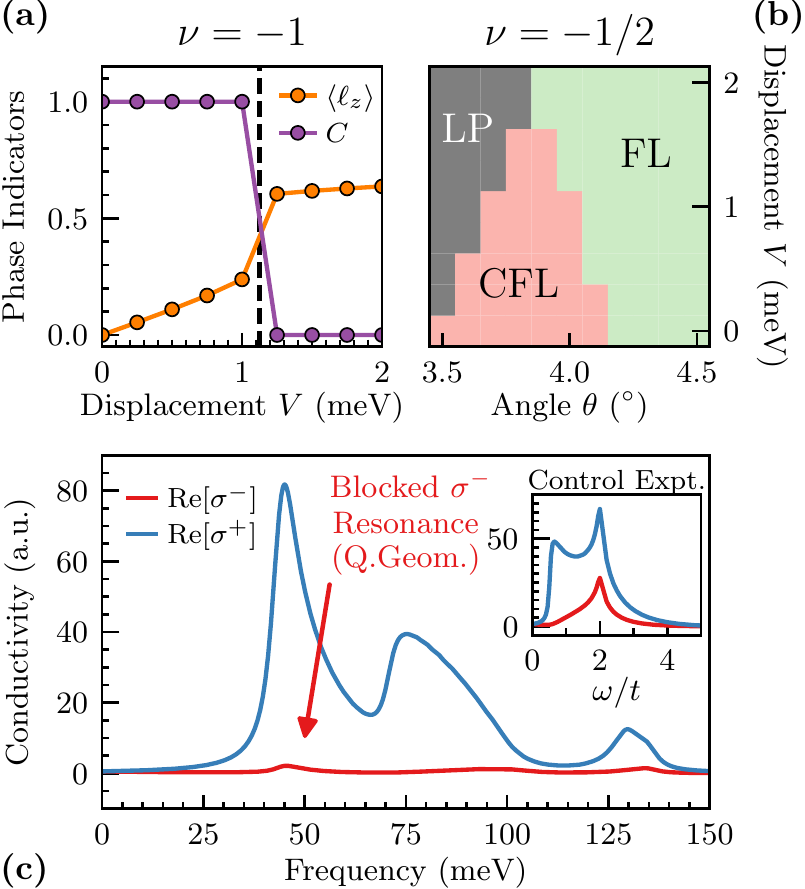}
    \caption{Many-body phase diagrams and optical responses. (a) Phase diagram at $\nu=-1$ with $\theta=3.7^\circ$ showing a transition from $C=1$ layer-unpolarized state to a $C=0$ layer polarized state.
    (b) Phase diagram of the topological regime at $\nu=-1/2$: The CFL phase is shown in red, whereas the green region corresponds to the FL phase. Here `LP' indicates a layer polarization instability determined from $\nu=-1$ SCHF.   
    (c) Direct optical probe of almost-ideal quantum geometry via an ``extinguished'' valence-valence optical responses in $\sigma^-$, Inset: the Haldane model at $(t,t_2)=(1,0.05)$ has non-ideal geometry. Parameters match Fig.~\ref{fig:bandstructure_geometry_optics}.}
    \label{fig:experimental_implications}
\end{figure}

\emph{Experimental Signatures.---}
We conclude with experimental implications of the quantum geometry and CFL phase.
Fig.~\ref{fig:experimental_implications}(a,b) show phase diagrams of t\ce{MoTe2}. At $\nu=-1$, SCHF finds the $\n{C}=1$ phase transitions to an valley and layer polarized phase at large $V$.
At $\nu=-1/2$, we find a broad CFL phase centered around $3.8^\circ$ that competes with layer polarized phases and $C=1$ Fermi liquids at larger $V$. The layer polarized region is estimated from SCHF at $\nu=-1$, where an interaction-driven layer-polarized state is more favorable. The phase diagram at $\nu=-3/4$ is similar (see SM), except the CFL is more sensitive to displacement field.

The almost ideal quantum geometry manifests optically. If a band with projector $P$ is vortexable, then $\hat{z}P = P\hat{z}P$ implies the velocity operator $\hat{v}^\pm = -i[\hat{x} \pm i \hat{y}, \hat{H}]$ must obey $(I-P) \hat{v}^+ P=0$, i.e. left-circularly polarized transitions are ``extinguished". This gives perfect circular dichroism:
\begin{equation}
    \frac{\sigma^+ - \sigma^-}{\sigma^+ + \sigma^-} = 1; \; \sigma^{\pm}(\omega) = \frac{i e^2 }{\hbar} \sum_{\substack{\v{k}, a \neq b=0}} \frac{f_{ab}}{\epsilon_{ab}} \frac{\n{\braket{\psi_{\v{k}a}|\hat{v}^{\pm}|\psi_{\v{k}b}}}^2}{\omega - \epsilon_{ab}}.
    \label{eq:perfect_dichroism}
\end{equation}
Here $\epsilon_{ab} = \epsilon_a - \epsilon_b$ are energy differences and $f_{ab} = f(\epsilon_a)-f(\epsilon_b)$ are Fermi factors. Fig.~\ref{fig:experimental_implications}(c) shows $\sigma^{\pm}$ for t\ce{MoTe2} at $\nu=-1$. As the $C=+1$ band is nearly vortexable, transitions from the second and third valence bands to the empty top valence band nearly vanish, giving nearly-perfect circular dichroism $>0.9$ at resonance. The inset shows a control experiment: the Haldane model has Chern bands $C = \pm 1$ but not ideal geometry; $\sigma^-$ is not extinguished there.

Finally, we discuss direct experimental probes of the zero-field CFL. While the CFL and the FL are both compressible and metallic, they differ in that the CFL's excitations have vanishing overlap with the electron in the limit of low energies, and CFs themselves are best thought of as (doubled) \textit{vortices} in the electronic fluid~\cite{HalperinLeeRead,Son,Metlitski_2016, WangSenthilCFL,WangSenthil}. This observation leads to a number of striking physical responses that differ strongly from Fermi liquids.  These include (i) a ``pseudogap" in the tunneling density of states $A(\omega)\propto e^{-\omega_0/\omega}$~\cite{STMCFLSong} as a function of bias $\omega$, which has been observed between two CFLs with a tunnel barrier~\cite{STMCFLEisenstein}; (ii) distinct DC conductivity in the clean limit: $\sigma_{xx} \rightarrow 0$ in a CFL in the absence of disorder $k_Fl\rightarrow \infty$, whereas in the FL, even in a Chern band, $\sigma_{xx}$ diverges~\cite{RevModPhys.82.1539}; (iii) strong violation of the Wiedemann-Franz law~\cite{WangSenthilCFL,WangSenthil} which compares heat and charge transport;
(iv) quantum oscillations with doping, that CFs feel a magnetic field $\propto (\nu - 1/2)$ and can fill Landau levels, leading to Jain-like~\cite{JainCF89} FCIs when fully developed, which can further be probed using geometric resonance with a one-dimensional periodic grating~\cite{HalperinLeeRead, WillettPRL,KangPRL}; (v) vanishing thermoelectric conductance $\alpha_{xx} = j_x/(-\partial_x T)$ due to approximate emergent particle-hole symmetry~\cite{CFLTransport,WangCooperHalperinStern}; (vi) surface acoustic wave attenuation, a contactless probe that measures $\sigma_{xx}(\v q) \propto |\v q|$ in the CFL~\cite{HalperinLeeRead}, as opposed to $\sigma_{xx} \propto |\v q|^{-1}$ in a clean FL~\cite{SAWWillett}.

Finally we highlight properties of zero field CFLs that transcend LLL physics. 
First, the Chern bands of \ce{MoTe2} have one effective magnetic flux quantum per moir\'e unit cell, translating to $\SI{160}{\tesla}$ at $3.7^\circ$. This vastly exceeds laboratory magnetic fields, leading to enhanced energy scales.
The lack of \emph{real} quantizing magnetic fields, however, opens up the possibility of employing zero field experimental probes such as high resolution angle-resolved photoemission spectroscopy (ARPES). Furthermore, the exponentially suppressed tunneling density of states of the CFL could be probed through tunneling from a proximate Fermi liquid state, or via spatial variation of the twist angle, which can be used to create a CFL-FL interface within the same sample. Our work does not rule out the possibility of a continuous quantum phase transition, driven by displacement field, between the CFL and FL~\cite{barkeshliContinuousTransitionsComposite2012}, which could be studied experimentally. Since the effective magnetic field of the TMD originates from spontaneous breaking of time reversal symmetry through valley polarization, rather than external magnetic field, domains between opposite valley polarizations and hence between time-reversal-related CFLs are expected. Transport properties across such a domain wall would interrogate composite fermions in an entirely new regime, and potentially shed light on their proposed Dirac character~\cite{Son,Metlitski_2016,WangSenthil}. Finally, we note that moir\'{e} phonons occur on the same scale as the effective magnetic length in this system; their interplay with CFL physics is unclear at present and worthy of future study.

\begin{acknowledgments}
We thank Junyeong Ahn, Ilya Esterlis, Eslam Khalaf, Jiaqi Cai, Richard Averitt, Darius Torchinsky, and especially Bertrand Halperin for helpful discussions. We acknowledge Michael Zaletel, Tomohiro Soejima, and Johannes Hauschild for ongoing and related collaborations. We sincerely thank Taige Wang for alerting us to the importance of layer polarization after our paper was announced on arXiv. Shortly after this manuscript was posted,~\cite{wang_magnon,goldman_acfl} appeared, which overlaps with parts of this work. Additionally, Ref.~\cite{onishi_optics} overlaps with the optical responses discussed here. Subsequent to our work, transport experiments~\cite{park2023observation} on twisted \ce{MoTe2} were performed and the results are consistent with our findings. A.V. is supported by the Simons Collaboration on Ultra-Quantum Matter, which is a grant from the Simons Foundation (651440, A.V.) and by the Center for Advancement of Topological Semimetals, an Energy Frontier Research Center funded by the US Department of Energy Office of Science, Office of Basic Energy Sciences, through the Ames Laboratory under contract No. DE-AC02-07CH11358. This research is
funded in part by the Gordon and Betty Moore Foundation’s EPiQS Initiative, Grant GBMF8683 to D.E.P.
\end{acknowledgments}
\bibliographystyle{unsrt}
\bibliography{ref.bib}

\begin{thebibliography}{100}
\bibitem{footnote2}
{The gaps computed here correspond to the collective neutral excitations. To
  get the activation gap of charged particles, one should generally consider
  $\Delta E=(E_{N+1}+E_{N-1}-E_N)/2$, where $E_N$ is the ground state energy of
  $N$ electrons.}
\bibitem{footnote4}
{Note that while continuous translation symmetry fixes $\psi^{\text{CF}}_{\v
  k}(\v r) = e^{i \v k \cdot \v r}$ in the LLL, different Bloch states may be
  used for systems with lattice translation symmetry. Likewise, the composite
  Fermi surface need not be flat.}
\end{thebibliography}

\appendix

\section{Continuum Model}
\begin{figure}
   \centering
   \includegraphics[width = 0.5\textwidth]{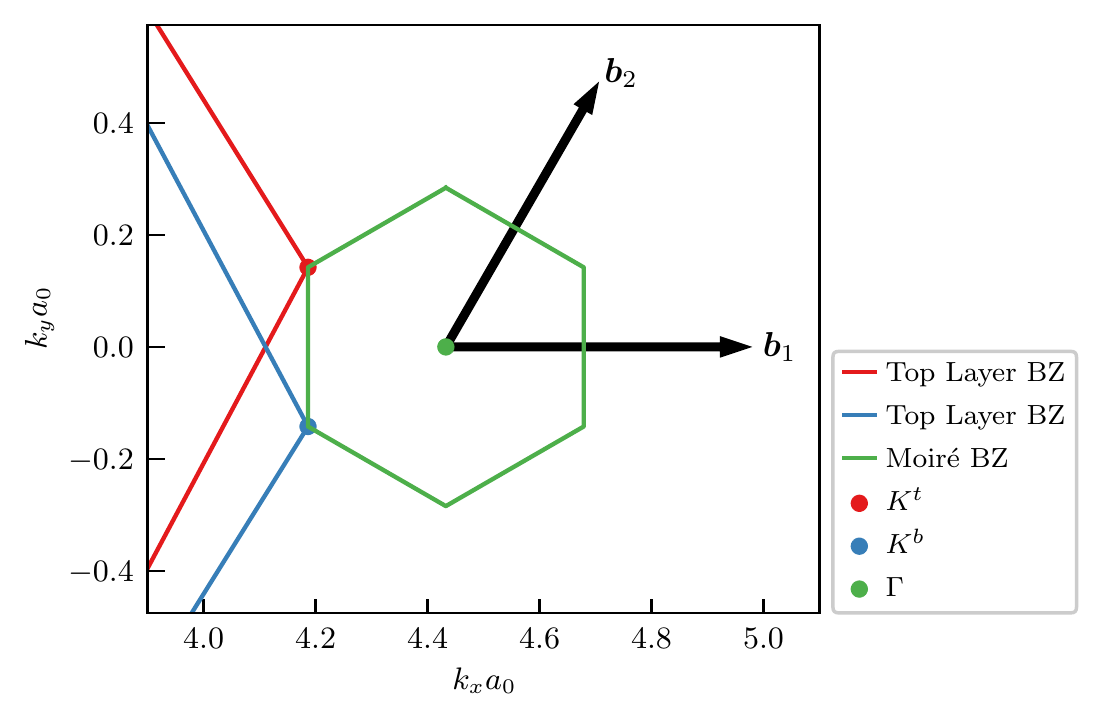}
   \caption{Geometry of the moir\'e Brillouin zone.}
   \label{fig:mBZ}
\end{figure}

In this Appendix we review the continuum model of TMDs~\cite{wu2019topological}. 

\textbf{Moir\'e Geometry}. We take monolayer TMD lattice vectors
\begin{equation}
   \bm R_{1,2} = a_0 \left(\frac{1}{2}, \mp\frac{\sqrt{3}}{2}\right),\qquad \bm G_{1,2} = \frac{2\pi}{a_0}\left(1,\mp\frac{1}{\sqrt{3}}\right)
\end{equation}
where $a_0=\SI{3.52}{\angstrom}$ is the monolayer TMD lattice constant~\cite{FCITMD23_Dixiao}. 

As usual, the moir\'e Brillouin zone for the $K$-valley has moir\'e lattice vectors
\begin{gather}
   \v{a}_{1} = \frac{a_0}{2\sin\frac{\theta}{2}}\left(\frac{\sqrt{3}}{2},-\frac{1}{2}\right), \v{a}_2 = \frac{a_0}{2\sin\frac{\theta}{2}}(0,1),\\ \v{b}_{1}=\frac{8\pi\sin \frac{\theta}{2}}{\sqrt{3}a_0}\left(1,0\right), \v{b}_j =R_{(j-1)\pi/3}\v{b}_1
\end{gather}
where $R_\theta$ is a rotation matrix. The geometry is shown in Fig.~\ref{fig:mBZ}, where one can see that two of the vertices of the hexagonal moir\'e Brillouin zone match the rotated $K$-points of the monolayer TMD:
\begin{eqnarray}
   \v{K}^\ell = R_{\ell\theta/2} \v{K}, \quad
   \v{K} = \frac{1}{3}\v{G}_1+\frac{1}{3}\v{G}_2
\end{eqnarray}
We often consider a commmensurate angle $\theta = 3.89^\circ$.

\begin{figure*}
\centering
\includegraphics[width=\linewidth]{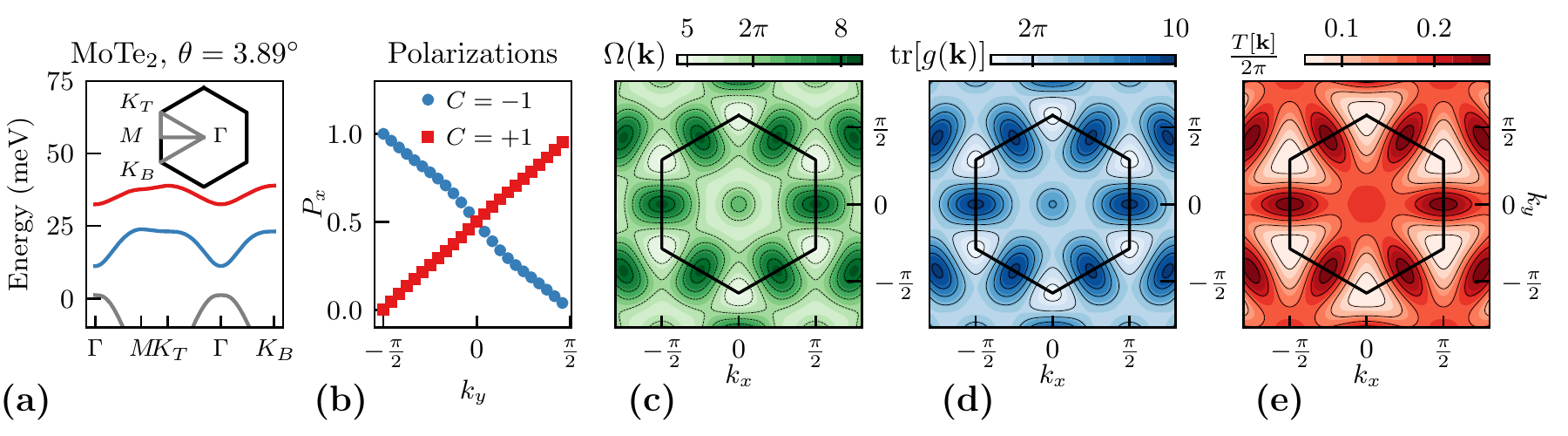}
\caption{(a) Band structure of twisted \ce{MoTe2} material at parameter $\theta=3.89^\circ$. The red band and blue bands are the top-most two hole bands. (b) Wilson line of the top-most two hole bands demonstrating they carry Chern number $C=-1$ and $C=+1$, respectively. (c) to (e): the momentum space quantum geometric properties of the top-most band (red), where (c) is the Berry curvature, (d) computes the trace of Fubini-Study metric and (e) depicts the deviation of trace condition in the Brillouin zone.}
\label{fig:app_bandstructure_quantum_geometry}
\end{figure*}

\textbf{Single-Particle Model}. The continuum model for a moir\'e bilayer TMD~\cite{wu2019topological} consists of four species of electrons, corresponding to two layers in each valley. Since the band gap from the valence band to the conduction band is large, the conduction bands are integrated out. Furthermore, due to spin-orbit coupling, the spin degree of freedom is locked to the valley degree of freedom: valence electrons at the $K$ valley have spin up, and valence electrons at the $K'$ valley have spin down. After these approximations, each valley has two degrees of freedom one for each layer. We focus on the $K$-valley and obtain the $K'$ valley by time-reversal, described below.

The single-particle Hamiltonian for holes in the $K$-valley is
\begin{equation}
   h_K = \begin{bmatrix} h^b(\v{r})+V/2 & T(\bm r) \\ T^\dag(\bm r) & h^t(\v{r})-V/2\end{bmatrix}
   \label{eq:hole_bandstructure_app}
\end{equation}
This Hamiltonian consists of three parts: the standard kinetic term, the moir\'e potentials and tunnelings and the displacement field. The kinetic part is
\begin{equation}
   h_{\text{kin}} = \begin{bmatrix} - \frac{(\v{p} - \hbar v_F \v{K}^b)^2}{2m^*} & 0 \\ 0 & - \frac{(\v{p} - \hbar v_F \v{K}^t)^2}{2m^*}\end{bmatrix}
\end{equation}
and the moir\'e potential and tunnelings are 
\begin{subequations}
\begin{align}
   h_{\text{moire}} &= \begin{bmatrix}
       \Delta^b(\v r) & T(\v r)\\
       T^\dagger(\v r) & \Delta^t(\v r)
   \end{bmatrix}\\
   \Delta^\ell(\v r) &= 2v\sum_{j=1,3,5}\cos(\v{b}_j\cdot\v{r}+\ell\psi),\\
   T(\v r)&=w(1+e^{i\v b_2\cdot \v r}+e^{i\v b_3\cdot \v r}).
   \end{align}
\end{subequations}
Fig. \ref{fig:app_bandstructure_quantum_geometry}(a) shows the bandstructure of Eq. \eqref{eq:hole_bandstructure_app} at $V=0$. As mentioned in the main text, we take parameters from recent first principles calculations of \ce{MoTe_2}~\cite{FCITMD23_Dixiao}: $(a_0, m^*, V,\psi,\omega) = (\SI{3.52}{\angstrom},0.6m_e, \SI{20.8}{\milli\electronvolt}, -107.7^\circ, \SI{-23.8}{\milli\electronvolt})$ at $\theta = 3.89^\circ$ (see also~\cite{FCITMD23_LF,wu2019topological,crepel2023chiral}). At these parameters, the top valence band for this model has Chern number $C=+1$, and the second-to-top band has Chern number $C=-1$.

\textbf{Symmetries.}
Some key symmetries of Eq.~\eqref{eq:hole_bandstructure_app} without displacement field are $C_3$, $C_{2y}$, and time-reversal $\mathcal{T}$. We give their action after applying a unitary transform to make their action simpler. Let $\tau = {K,K'} = {1,-1}$ index valleys and define
\begin{equation}    
   h'_{\tau}(\bm r) = U_{\tau}(\bm r)h_{\tau}(\bm r) U_\tau^\dagger(\bm r), U_\tau(\bm r) = \begin{bmatrix}
   e^{-i\tau {\bm K}^b\cdot{\bm r}} & 0\\
   0 & e^{-i\tau {\bm K}^t\cdot{\bm r}}
   \end{bmatrix}.
\end{equation}
Then symmetries then act on $h'$ as follows:
\begin{subequations}
\begin{align}
   C_3 h'_{\tau} (\bm r)C_3^{-1} &= h'_{\tau}(e^{\frac{2\pi i}{3}\ell_y}\bm r),\\
   C_{2y}h'_{\tau}(x,y)(C_{2y})^{-1} &= \ell_x h'_{-\tau}(-x,y)\ell_x,\\
   {\mathcal{T}}h'_{\tau}(x,y){\mathcal{T}}^{-1} &= h'_{-\tau}(x,y)^*,
\end{align}
   \label{eq:symmetries}
\end{subequations}
where we use layer-space Pauli matrices $\ell_{0,x,y,z}$.

\textbf{Interactions}. Consider creation operators
\begin{equation}
   \hat{c}^\dagger_{\v{k} \tau b} \ket{0} = \ket{\psi_{\v{k} \tau b}}, \quad \hat{h}_{\tau}(\v{k}) \ket{\psi_{\v{k} \tau b}} = \epsilon_{\v{k} \tau b} \ket{\psi_{\v{k} \tau b}}
\end{equation}
where $\tau \in \st{K,K'}$ and $b$ labels bands. We will often use $\v{c}^\dagger_{\v{k}}$ as a vector in (band,valley) for concision. As mentioned in the main text we consider an interacting Hamiltonian 
\begin{equation}
\hat{H} = -\hat{h} + \frac{1}{2A}\sum_{\v{q}} V_{\v{q}} :\hat{\rho}_{\v{q}} \hat{\rho}_{-\v{q}}:,\,\,\, V_{q} = \frac{2\pi \tanh(q d)}{\epsilon_r \epsilon_0 q},
\label{eq:app_TMD_hamiltonian}
\end{equation}
where
\begin{equation}
\hat{\rho}_{\v q} = \sum_{\v{k}} \hat{\v{c}}_{\v{k}}^\dagger \Lambda_{\v{q}}(\v{k}) \hat{\v{c}}_{\v{k}}; \quad \Lambda_{\v{q}}(\v{k})  = \braket{u_{\v{k}}|u_{\v{k}+\v{q}}}
\label{eq:app_density_operator}
\end{equation}
is the density operator, $A$ is sample area,  normal ordering is relative to filling $\nu=0$, and $\epsilon_r \approx 5-15$ is the dielectric constant. As usual, $V_{\v{q}}$ corresponds to double-gate screened Coulomb interactions with gates at distance $d=\SIrange{100}{300}{\angstrom}$. Here interactions are added directly to the ``bare'' kinetic term $\hat{h} = \sum_{\v{k}} \sum_{\tau \in \st{K,K'}} \hat{c}_{\v{k}\tau}^\dagger h_\tau(\v{k}) \hat{c}_{\v{k}\tau}$, where $h_K$, Eq. \eqref{eq:hole_bandstructure_app}, is determined by, e.g., first principles calculations of the insulating state where the chemical potential is inside the large valence-conduction gap. This is markedly different from interacting gapless systems like twisted bilayer graphene where the single-particle Hamiltonian is quite different from the bare Hamiltonian, and one must apply ``Hartree-Fock subtraction''~\cite{Zaletel_PRX20,parker2021field} or other renormalization procedures~\cite{Oscar_hiddensym} to determine the bare kinetic term. Furthermore, in this situation, integrating out all but the top few valence bands at mean-field level is equivalent to \textit{projecting} to those top few bands; we may form a low-energy effective model of the top valence band by simply restricting the sum over bands in Eqs. \eqref{eq:app_TMD_hamiltonian}, \eqref{eq:app_density_operator}.

\textbf{Self-Consistent Hartree-Fock} In the main text and Fig. 1 we make use of standard self-consistent Hartree-Fock (SCHF) calculations. We briefly detail them here for completeness. Consider a single-particle density matrix 
$P(\v{k})_{ab} = \braket{c_{\v{k}b} c^\dagger_{\v{k}a}}$. Such density matrices are bijective to Slater determinant states. As usual, we define Hartree and Fock Hamiltonians
\begin{subequations}
\begin{align}
   h_H[P](\v{k}) &= \frac{1}{A} \sum_{\v{g}} V_{\v{g}} \Lambda_{\v{G}}(\v{k}) \sum_{\v{k}} \operatorname{tr}[P(\v{k}) \Lambda_{\v{G}}^\dagger(\v{k})]\\
   h_F[P](\v{k}) &= -\frac{1}{A} \sum_{\v{q}} V_{\v{q}} \Lambda_{\v{q}}(\v{k}) P([\v{k}+\v{q}]) \Lambda_{\v{q}}(\v{k})^\dagger.
\end{align}\label{eq:Hartree_Fock_equations}\end{subequations}
Here form factors $\Lambda_{\v{q}}(\v{k})$ are matrices in (band,valley) space. In terms of these, the energy of a Slater determinant state has energy $E[P] = \braket{\hat{H}}_P = \frac{1}{2}\operatorname{Tr}[P(h + h_H[P]+h_F[P])]$ where $h$ is the single-particle Hamiltonian and $\operatorname{Tr}$ is the trace over both $\v{k}$ and bands. If the self-consistency condition $[P, h_{HF}[P]] = 0$ is satisfied for all $\v{k}$, then $E[P]$ is a local minimum. In practice we consider the top four valence bands of both valleys on a $k$-grid of size $24 \times 24$, and use the standard ``ODA" algorithm to achieve the self-consistency condition at $\nu=-1$ to numerical precision,  imposing valley polarization and $C_3$ symmetry. Fig. 1 shows the resulting SCHF bandstructure along high-symmetry lines.

\section{Quantum Geometry}
In this section we review momentum-space band geometry~\cite{Parameswaran_2013} and its relationship to the real-space vortexability criterion~\cite{LedwithVishwanathParker22}. We begin by introducing
\begin{gather}
   \eta^{\mu\nu}(\bm k) = \sum_{a}\bra{\partial_{k_\mu}u_{\bm k a}}Q(\bm k)\ket{\partial_{k_\nu}u_{\bm ka}},\\
   Q(\bm k) = 1-\sum_{a}\ket{u_{\bm ka}}\bra{u_{\bm ka}};~P = \sum_{\v{k},a}\ket{\psi_{\bm ka}}\bra{\psi_{\bm ka}},
\end{gather}
where $\eta^{\mu\nu}$ is the so-called quantum geometric tensor. The real and imaginary part of the quantum geometric tensor defines the Fubini-Study metric and Berry curvature, respectively:
\begin{equation}
   g^{\mu\nu}(\bm k)=\textrm{Re}(\eta^{\mu\nu}),\quad \Omega(\bm k)\epsilon^{\mu\nu}=-\frac{1}{2}\textrm{Im}(\eta^{\mu\nu}).
\end{equation}

The metric $g^{\mu\nu}$ and the Berry curvature two form $\Omega(\bm k)\epsilon^{\mu\nu}$ assign the Brillouin zone a Riemann structure and symplectic structure, respectively. When these two structures are ``compatible'', holomorphic coordinates can be defined~\cite{kahlerband1,kahlerband2,kahlerband3}. 

A sufficient compatibility condition is given by $\Tr g(\bm k)=\Omega(\bm k)$, which we focus on here, though we briefly comment on possible generalizations.
Refs.~\cite{kahlerband1,kahlerband2,kahlerband3} discuss the most general momentum-space compatibility condition (the so-called ``determinant condition"~\cite{Parameswaran_2013,Jackson:2015aa}. Ref.~\cite{JieWang_exactlldescription} focuses a sub-class related to the trace condition by linear transformations; only this sub-class implies real-space vortexability, and an orthogonal argument based on ``forgetting translation symmetry" further distinguishes this class~\cite{LedwithVishwanathParker22}. Vortexability may also be related to a K\"{a}hler geometry in real-space, and different choices of vortex functions $z \neq x+iy$ correspond to a wide variety of different real-space K\"{a}hler structures, most of which do not correspond to ideal momentum space geometry in the traditional sense~\cite{LedwithVishwanathParker22}.

The following conditions are equivalent conditions~\cite{LedwithVishwanathParker22,JieWang_exactlldescription,kahlerband1,kahlerband2,kahlerband3}:
\begin{enumerate}
   \item{Trace Condition}: $\Tr g(\bm k)=\Omega(\bm k)$, for all momentum $\bm k$ in the first Brillouin zone. This condition is equivalent to the saturation of the general trace inequality: $\Tr g(\bm k)\geq\Omega(\bm k)$.
   \item{Momentum Space Holomorphicity}: $Q(\bm k)\ket{\overline{\partial}_k u_{\bm k a}} = 0$, where $\overline{\partial}_k = \partial_{k_x}+i\partial_{k_y}$, for all momentum $\bm k$ in the first Brillouin zone. This condition is equivalent to the existence of an unnormalized gauge in momentum space for the Bloch wavefunctions $\ket{u_{\v{k} a}}$ such that $u_{\v{k} a}$ is holomorphic in $k_x+ik_y$.
   \item{Vortexability}: $\hat{z}\psi = P \hat{z}\psi$, where $z=x+iy$, for all states $\psi$ belonging to the bands in the projector $P$. This condition is equivalent to the one used in text $\hat{z}P=P\hat{z}P$.
\end{enumerate}

In Fig.~\ref{fig:app_bandstructure_quantum_geometry}(c, d, e) we show the quantum geometry of the top-most valence band. Particularly, it almost saturates the trace bound $T=\int d^2\v{k} \left(\Tr g(\bm k)-\Omega(\bm k)\right)\geq 0$.

The momentum space holomorphicity condition is extremely powerful for constraining the form of flat band wavefunctions, and has been used to prove the generality of the wavefunctions Eq. (3) for single $C=1$ bands and related results for single bands with $C>1$~\cite{JieWang_exactlldescription,JW_origin_22,junkaidonghighC22}.

\section{Perfect Circular Dichroism}

We now outline how to to derive the circular dichroism result Eq. 5. Consider a band (or set of bands) with projector $P~=~\sum_{\v{k},b} \ket{\psi_{\v{k}b}} \bra{\psi_{\v{k}b}}$ and define an orthogonal projector $Q = I - P.$ The vortexability condition $\hat{z}P=P\hat{z}P$ can be rewritten as $Q\hat{z}P=0$ or, taking the Hermitian conjugate, $P\hat{z}^\dagger Q=0$. Therefore
\begin{equation}
0 = ||P\hat{z}^\dagger Q|| = \sum_{a \in P, b \in Q} \sum_{\v{k}} \n{\braket{\psi_{\v{k}a}|\hat{z}^\dagger |\psi_{\v{k}b}}}^2.
\end{equation}
Since each term is non-negative, they must all vanish, implying
\begin{equation}
   \braket{\psi_{\v{k}a}|\hat{z}^\dagger |\psi_{\v{k}b}} = 0  \text{ for } a \in P, b \in Q.
   \label{eq:app_vortexability_position_matrix_elements}
\end{equation}
We consider optical response in the velocity gauge, which involve matrix elements 
\begin{equation}
   v^-_{ab}=\braket{\psi_{\v{k}a}|\hat{v}^-|\psi_{\v{k}b}} 
\end{equation}
of the velocity operator $\hat{v}^\mu = -i[\hat{r}^\mu, \hat{H}]$.
Since each both $\ket{\psi_{\v{k}a}}$ and $\ket{\psi_{\v{k}b}}$ are eigenstates, Eq. \eqref{eq:app_vortexability_position_matrix_elements} implies
\begin{eqnarray}
   v^-_{ab}=-i(\epsilon_{\v{k}b}-\epsilon_{\v{k}a})\braket{\psi_{\v{k}a}|\hat{z}^\dagger |\psi_{\v{k}b}}=0.
\end{eqnarray}

The linear optical response for interband transitions from $Q$ to $P$ under circularly polarized light is given by the Kubo formula
\begin{eqnarray}
   \sigma^{\pm}_{Q\to P}(\omega) = \frac{i e^2 }{\hbar} \sum_{\substack{\v{k}, a\in Q, b\in P}} \frac{f_{ab}}{\epsilon_{ab}} \frac{\n{\braket{\psi_{\v{k}a}|\hat{v}^{\pm}|\psi_{\v{k}b}}}^2}{\omega - \epsilon_{ab}}.
\end{eqnarray}
If the chemical potential is placed such that some of the states corresponding to $Q$ are filled and $P$ is unfilled, then the right handed optical transitions corresponding to $Q\to P$ will be extinguished: $\sigma^-_{Q\to P}(\omega) =0$ uniformly, whereas $\sigma^+$ is generically non-zero. 

Observing this effect --- which reflects the quality of the quantum geometry rather than topology --- likely requires a large sample of high quality (uniform twist angle, low strain, etc), but does not require low temperature. We note that SCHF typically overestimates the gap at $\nu=-1$, since it neglects, for instance, the effects of disorder.

\section{Parton Theory of the Zero-Field CFL}

In this section we discuss a parton-based approach, first introduced by Refs.~\cite{Jain1989incompressible,wen1991nonabelian}, that elucidates how a charge neutral composite fermion that feels zero magnetic flux can appear at low energy despite zero magnetic field. We note that the two most prominent composite fermion theories of flux-feeling electrons, those of HLR and Son, rely on beginning with a flux-feeling particle. Indeed, the approach of HLR~\cite{HalperinLeeRead} attaches emergent gauge flux to the electron to cancel magnetic flux, while the particle-vortex duality of Son\cite{Son} relates the density of composite fermions to the magnetic field felt by electrons. Naively applied to the zero-field TMD, these approaches would lead to composite fermions that either feel flux or have zero density, respectively, neither of which is not suitable for the phenomenology of the CFL. While other parton-based approaches\cite{luSymmetryprotectedFractionalChern2012a,mcgreevyWaveFunctionsFractional2012,murthyCompositeFermionsFractionally2011,murthyHamiltonianTheoryFractionally2012,barkeshliContinuousTransitionsComposite2012,barkeshliContinuousTransitionFractional2014} have been developed to discuss composite fermions in Chern bands, to our knowledge none enable \emph{any} Landau level theory to be applied.

We pause to comment that, despite the presence of a zero field Chern band, the TMD electrons feel zero flux. While this is clearly true from the perspective of the kinetic Hamiltonian \eqref{eq:app_TMD_hamiltonian}, we will see that the zero average flux felt is a property of the low energy Chern band wavefunctions as well. It is easiest to make this notion precise when there is discrete translation symmetry. In magnetic systems, where Hamiltonians depend on derivatives through the covariant derivative, $-i \partial_\mu - a_\mu(\v r)$, translation operators come with position-dependent ``twists"
\begin{equation}
   T_{\v R } \psi(\v r) = e^{-i \varphi(\v r)} \psi(\v r + \v R), \,\, \partial_\mu \varphi_{\v R}(\v r) = a_\mu(\v r + \v R) - a_\mu(\v r)
   \quad  
\end{equation}
such that the commutator 
\begin{equation}
T_{\v R_1} T_{\v R_2}  = e^{i \Phi}  T_{\v R_2} T_{\v R_1}
\end{equation}
is related to the flux enclosed
\begin{equation}
\begin{aligned}
   \Phi  & = \varphi_{\v R_1}(\v r + \v R_2) - \varphi_{\v R_1}(\v r) + \varphi_{\v R_2}(\v r) - \varphi_{\v R_2}(\v r + \v R_1) \\
    & = \oint_{\partial \text{UC}} \v a \cdot \v d \v r = \int_{\text{UC}} \v \nabla \times \v a \,\, d^2 \v r.
    \end{aligned}
   \end{equation}
Here the line integration is taken over the boundary of the parallelogram unit cell spanned by the vectors $\v R_1$ and $\v R_2$ and the surface integration is over the area of the parallelogram.

Let us suppose that $\Phi$ is an integer times $2\pi$, such that $[T_{\v R_1} ,T_{\v R_2}] = 0$ and we can find Bloch states such that $T_{\v R_{1,2}} \phi_{\v k} = e^{i \v k \cdot \v r} \phi_{\v k}$. Then we can identify the average flux per unit cell as the winding of the phase of the Bloch state around the unit cell:
\begin{equation}
\begin{aligned}
 & \oint \v{\nabla}  \text{Im} \log \psi_{\v k l}(\v r) \cdot d \v r  \\
 & = \varphi_{\v R_1}(\v r + \v R_2) - \varphi_{\v R_1}(\v r) + \varphi_{\v R_2}(\v r) - \varphi_{\v R_2}(\v r + \v R_1) \\
 & = \Phi(\v R_1, \v R_2) = \int_{\text{UC}} \v \nabla \times \v a \, d^2 \v{r}
 \end{aligned}
 \label{eq:fluxdefn}
\end{equation}
We emphasize that the above result implies that the average flux is meaningful even without reference to a UV Hamiltonian; it measures the phase winding per unit cell of any translationally symmetric wavefunction. While particles in the LLL feel an average flux, particles in the twisted TMD feel zero flux, even those with wavefunctions in the low-energy Chern band.

To remedy this situation, and make LLL theories of CFLs accessible, we will fractionalize the TMD electron on layer $l$ as $\psi_l = \phi \chi_l$, where $\phi$ carries the electric charge, and $\chi$ is a bosonic ``layeron'', a layer-space analog of a spinon, that satisfies $\sum_l |\chi_l(\v{r})|^2 = 1$ for all $\v{r}$. Associated to this decomposition, there is a $U(1)$ gauge redundancy and an associated emergent gauge field $\alpha_\mu$ for which $\phi$ has charge $1$ and $\chi$ has charge $-1$. We will choose $\langle \v{\nabla} \times \v \alpha \rangle = B_{\text{eff}} = 2 \pi/A_{\text{uc}}$, so that $\phi$ now feels a net flux of one flux quantum per unit cell. We will condense the bosonic layerons by choosing them to all be in the same magnetic-translationally-symmetric state $\zeta_l(\v{r})$; while this can be generalized to any superfluid wavefunction, we will see that this simple ``Bose Einstein condensate" form will match well with the vortexable limit of the TMD Chern band wavefunctions.

With this parton ansatz, the Gutzwiller projected wavefunctions always have the form
\begin{equation}
 \Psi_{\{ l_i \}}( \{ \v r_i \})  = \Psi_\phi \Psi \chi = \Psi_\phi( \v r_i) \prod_i \zeta_{l_i}(\v r_i),
 \label{eq:reducetoLLL}
\end{equation}
where $\Psi_\phi$ is a wavefunction of particles that feel one flux quantum per unit cell. We may now apply any preferred LLL theory to the $\phi$ particles to generate choices for $\Psi_\phi$. Relatedly, the low energy effective actions all have the form
\begin{equation}
   L = L_{\text{LLL}}(\phi, A_\mu + \alpha_\mu)
   \label{eq:generaleffective_lagrangian}
\end{equation}
such that $\alpha_\mu$ precisely makes up for the lack of a background magnetic field. This addresses conceptual issues with the HLR\cite{HalperinLeeRead} and Son\cite{Son} theories in Chern bands: flux attachment and particle-vortex duality, respectively, should be applied to the parton $\phi$, which feels flux, not the electron $\psi$.

The parton ansatz $\psi_l = \phi \chi_l$ is \textit{exact} in the vortexable limit. To see this, let us recall the wavefunctions of $C=1$ vortexable bands, reviewed in the main text and copied below for convenience:
\begin{equation}
   \psi_{\v k l}(\v r) = \phi_{\v k}(\v r) \zeta_l(\v r).
   \label{eq:generalform_singleparticle_supp}
\end{equation}
Here, $\phi_{\v k}(\v r)$ is a wavefunction of a Dirac particle in an inhomogeneous magnetic field, with one flux quanta per unit cell, and $\zeta_l(\v r)$ is a layer-space vector that is $\v k$-independent. The wavefunctions $\phi$ and $\zeta$ are symmetric under \emph{magnetic} translations with opposite magnetic twists. Thus, implicit in \eqref{eq:generalform_singleparticle_supp} is the same gauge redundancy and charge-layer separation reviewed above. Furthermore, the $\v k$-independence of $\zeta$ implies that \emph{all} many-body wavefunctions associated to the band \eqref{eq:generalform_singleparticle_supp} have the form \eqref{eq:reducetoLLL}.

From a more physical perspective, the parton ansatz $\psi_l = \phi \chi_l$ may also be motivated by the well-known skyrmionic character of the layer index in the TMD band (general amongst all vortexable bands) and the ``effective field" it generates\cite{wu2019topological,YaoTMD1}. 
Indeed the tunneling in the TMD Hamiltonian of interest, $\v \Delta(\v{r}) \cdot \v \sigma$, has a skyrmionic pattern in the unit cell:
\begin{equation}
 \frac{1}{4\pi} \int_\text{UC}   \v n_{\Delta} \cdot (\partial_x \v n_\Delta \times \partial_y \v n_\Delta) = -1,
 \label{eq:skyrmcharge}
\end{equation}
where $\v n_\Delta(\v{r}) = \v \Delta(\v{r})/\n{\v \Delta(\v{r})}$
Outside the vortexable limit, this skyrmionic texture can still be thought of as generating an effective magnetic field in a sense that we now describe. The wavefunctions can still be decomposed as 
\begin{equation}
   \psi_{\v{k}}(\v{r}) = \phi_{\v{k}}(\v{r}) \chi_{\v{k} l}(\v{r})
\end{equation}
where now $\chi$ is $\v{k}$-depndent. The winding of $\chi$, computed from $\v{n}_\chi = \chi^\dag \v{\sigma} \chi$ at each $\v{k}$, is expected to be the same, either by continuity from the vortexable limit, or since it is energetically favorable for it to follow the direction of the tunneling matrix, at least roughly. Note, however, that the skyrmion charge is equal to the number of flux quanta of the gauge field $a^{\chi}_\mu = -i \chi^\dag \partial_\mu \chi$ through the unit cell:
\begin{equation}
   \frac{1}{4\pi} \int_\text{UC}   \v n_{\chi} \cdot (\partial_x \v n_\chi \times \partial_y \v n_\chi) = \frac{1}{2\pi}  \int_{\text{UC}} \v \nabla \times \v a^\chi d^2 \v{r}.
\end{equation}
Furthermore it is straightforward to show that the net flux of $a^\chi$ corresponds to a net flux felt by $\chi$ in the sense of \eqref{eq:fluxdefn}: a nonzero phase winding per unit cell. Indeed, if $\chi_l(\v{r}) = e^{-i \varphi(\v{r})}\chi_l(\v{r} + \v{R}_{1,2})$ then we have
\begin{equation}
\begin{aligned}
 & \oint \v{\nabla}  \text{Im} \log \chi_{l}(\v r) \cdot d \v r  \\
 & = \varphi_{\v R_1}(\v r + \v R_2) - \varphi_{\v R_1}(\v r) + \varphi_{\v R_2}(\v r) - \varphi_{\v R_2}(\v r + \v R_1) \\
 & = \int_{\text{UC}} \v \nabla \times \v a^\chi \, d^2 \v{r}
 \end{aligned}
 \label{eq:fluxdefn2}
\end{equation}
where we used the boundary conditions of $\chi$ to evaluate both integrals, obtaining the middle expression in both cases. 

We therefore conclude that the if $\chi_l$ has a skyrmion texture, its wavefunctions must have corresponding a winding per unit cell as if $\chi_l$ coupled to a gauge field with an average flux of $-2\pi$ per unit cell. The corresponding parton $\phi$ must therefore feel $+2\pi$ flux, again leading to the parton ansatz described above.

\section{Layer Polarization versus Band Projection}
\label{sec:app_band_projection}

\begin{figure*}
    \centering
    \includegraphics[width=0.75\textwidth]{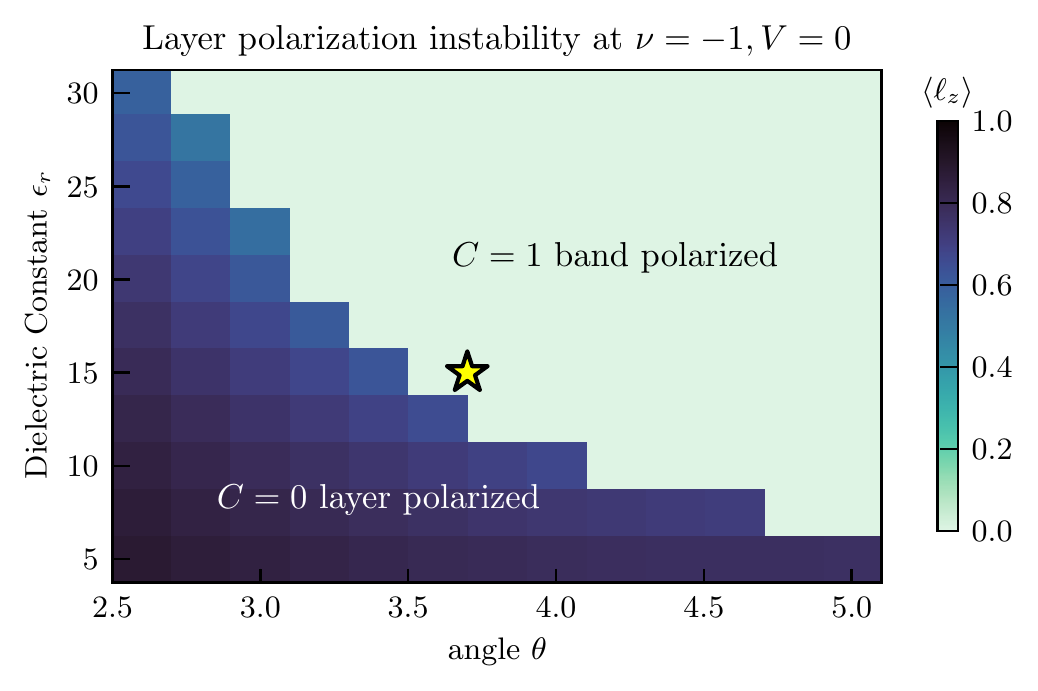}
    \caption{SCHF phase diagram at $\nu=-1$ in the $(\theta,\epsilon_r)$ plane, showing an interaction-induced layer polarized phase in the dark region. The band projection approximation is valid at $\nu=-1$ in the light region. The standard parameters used in the main text, $\theta=3.7^\circ, \epsilon_r=15$, are marked by a yellow star. The SCHF are performed on a $24\times 24$ torus, where we take the top two valence bands for each valley.}
    \label{fig:HFLP}
\end{figure*}

In this Appendix, we discuss a subtle issue arising from the interplay of layer polarization versus band projection. To carry out many-body numerics, it is always preferable to work within the smallest possible Hilbert space, i.e. the Hilbert space of a single flat band. Such an approximation is valid only when the gap to other bands is large, preventing band mixing. For \ce{MoTe2}, however, there is significant band mixing in a significant range of parameter space. Below we identify the regime where band-projection is permissible. We then discuss how band projected numerics in the `impermissible' regime can nevertheless be used as a rigorous numerical tool for phase identification via an adiabatic connection.

\subsection{Layer Polarization Physics}

We start at filling $\nu=-1$, where an instability to a layer-polarized (LP) phase competes with the quantum anomalous hall (QAH) phase, even a at displacement field $V=0$. This instability was previously pointed out in Abouelkomsan \textit{et al.}~\cite{abouelkomsan2022multiferroicity} and Wang \textit{et al.}~\cite{wang_magnon}. In this phase, the top two bands \textit{in the same valley} are strongly mixed, yielding a topologically trivial insulator that remains valley polarized. Fig. \ref{fig:HFLP} shows the phase diagram at $\nu=-1$ within self-consistent Hartree-Fock (SCHF).

 Intuitively, the layer-polarized phase enables electrons to move further apart, thus minimizing the Coulomb energy. On the top \ce{MoTe2} layer, the electron densities are concentrated on the MX sites, whereas on the bottom layer, the electron densities are concentrated on the XM sites \cite{YaoTMD1}. In the quantum anomalous Hall phase, the unoccupied band has densities on both layers and both sites, whereas in the layer polarized phase, the unoccupied band has densities only on on layer and one site. If the interaction is strong enough, the layer-unpolarized quantum anomalous hall phase will cost more energy, since the MX-XM distance is shorter than the MX-MX distance. This explains the appearance of the layer-polarized phase at strong interactions (small $\epsilon_r$) in Fig. \ref{fig:HFLP}. At slightly weaker interactions, where the interaction strength is not much larger than the bandgap, one expects the quantum anomalous Hall phase to be more energetically favorable. We note that the phase boundary here depends rather strongly on microscopic parameters. Their values are drawn from DFT calculations that, especially for $\psi$, may not be representative of experimental values.

\subsection{Regime of Flat Band Projection}

Given this instability to band mixing and layer-polarized physics, when is a band-projected model valid? Were numerical resources not an issue, one could perform calculations in an expanded two-band Hilbert space that included band-mixing exactly. In the present context, however, we expect gapless ground states that require large system sizes to stablize and identify.  As a practical matter, this restricts us to either single-band Hilbert spaces or extremely coarse momentum resolution with larger numbers of bands in exact diagonalization. We are therefore motivated to restrict to the flat band Hilbert space.

We make the assumption that, whenever the $\nu=-1$ state is in the QAH phase, then a band-projected model will faithfully capture the top valence band at partial fillings. This is reasonable since the top valence band is only slightly mixed with the lower bands in SCHF, meaning that the relevant Hilbert space at partial fillings should be that of the top band. Confirming this assumption within large-scale multi-band numerics is an important problem for future work. Hence, to properly identify the CFL in many-body numerics, we have to perform the band projection approximation at $\nu=-1/2$. We note this assumption is standard within the literature on \ce{MoTe2} (see e.g. ~\cite{Kaisun_FCI21, FCITMD23_Dixiao,FCITMD23_LF,goldman_acfl}), as well as twisted bilayer graphene where the relevant gaps are somewhat larger. We apply this assumption at parameters $(\theta,\epsilon_r) = (3.7^\circ, 15)$ in the main text, where the system is within the QAH phase at $\nu=-1$. Our expectation is that band projected numerics at these parameters faithfully represent the universal features of the CFL phase that should appear in experiments.

\subsection{Adiabatic Connection} 

Even though ground states in the flat-band-projected Hilbert space are not global ground states at small $\epsilon_r$, we can still use these states as a \textit{numerical tool} to understand the phases present. This is of immense practical utility, because the strongly-correlated CFL phase is most easily recognized in the limit of large interactions. To recognize it numerically, we may adopt the following procedure: (1) show a CFL is present in the band-projected model at small $\epsilon_r$, (2) show that the ground state in the band-projected model at a larger $\epsilon_r$ is adiabatically connected to the first point, (3) by our assumption above, the band-projected ground is adiabatically connected to the ground state in the unrestricted Hilbert space. We could, of course, have simply started from step (2), but step (1) is much deeper in the CFL. That point is therefore is easier to converge, and numerically cheaper to study in detail.

\begin{figure*}
    \centering
    \includegraphics[width=\textwidth]{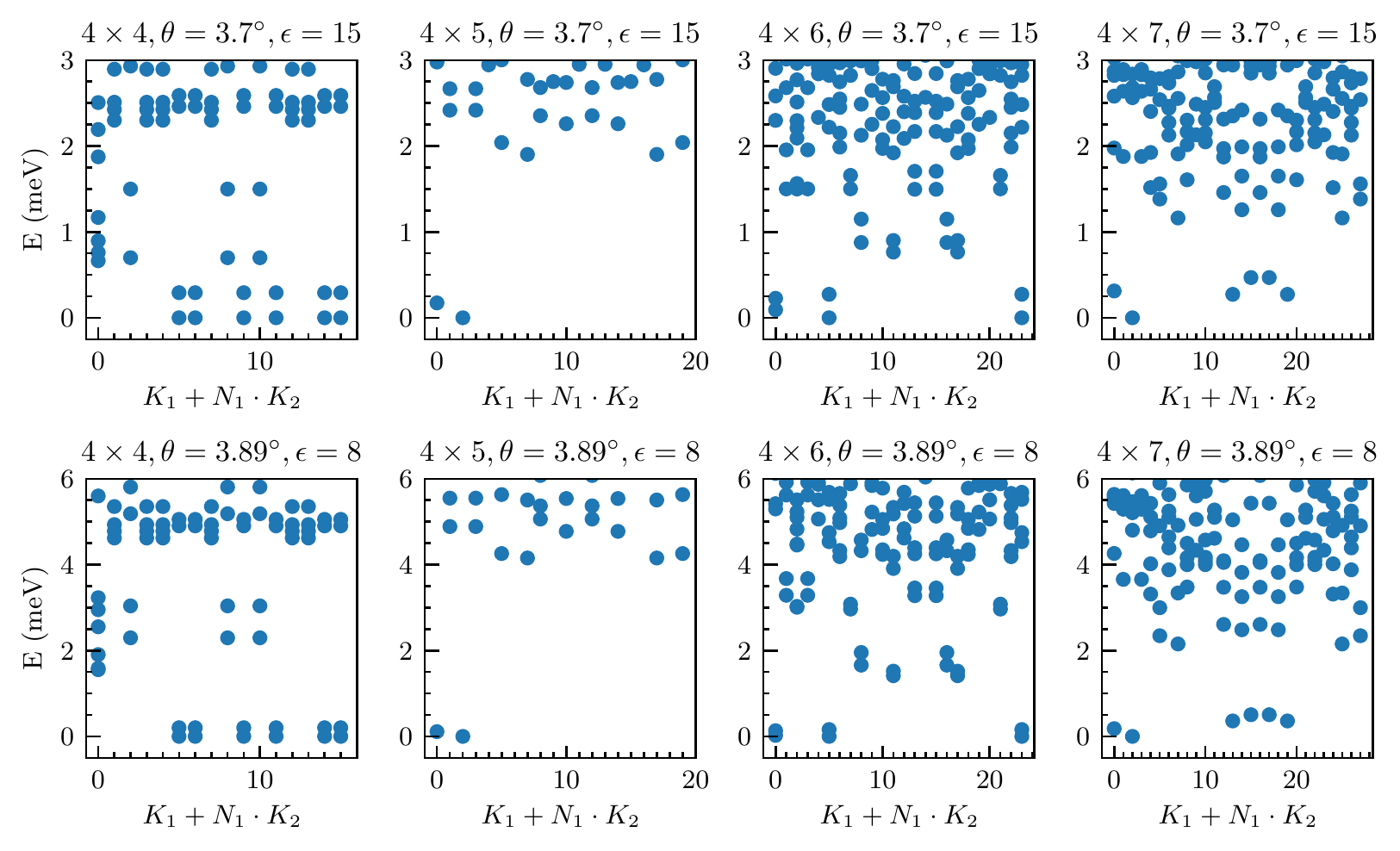}
    \caption{Comparison of ED spectra at many system sizes within the band projection approximation. The first row uses parameters $(V,\epsilon_r,\theta)=(0,15,3.7^\circ)$ and the second row uses parameters $(V,\epsilon_r,\theta)=(0,8,3.89^\circ)$. One can construct a one-to-one mapping between the low-energy spectra for these two sets of parameters.}
    \label{fig:EDparamchange}
\end{figure*}

\begin{figure*}
    \centering
    \includegraphics[width=0.5\textwidth]{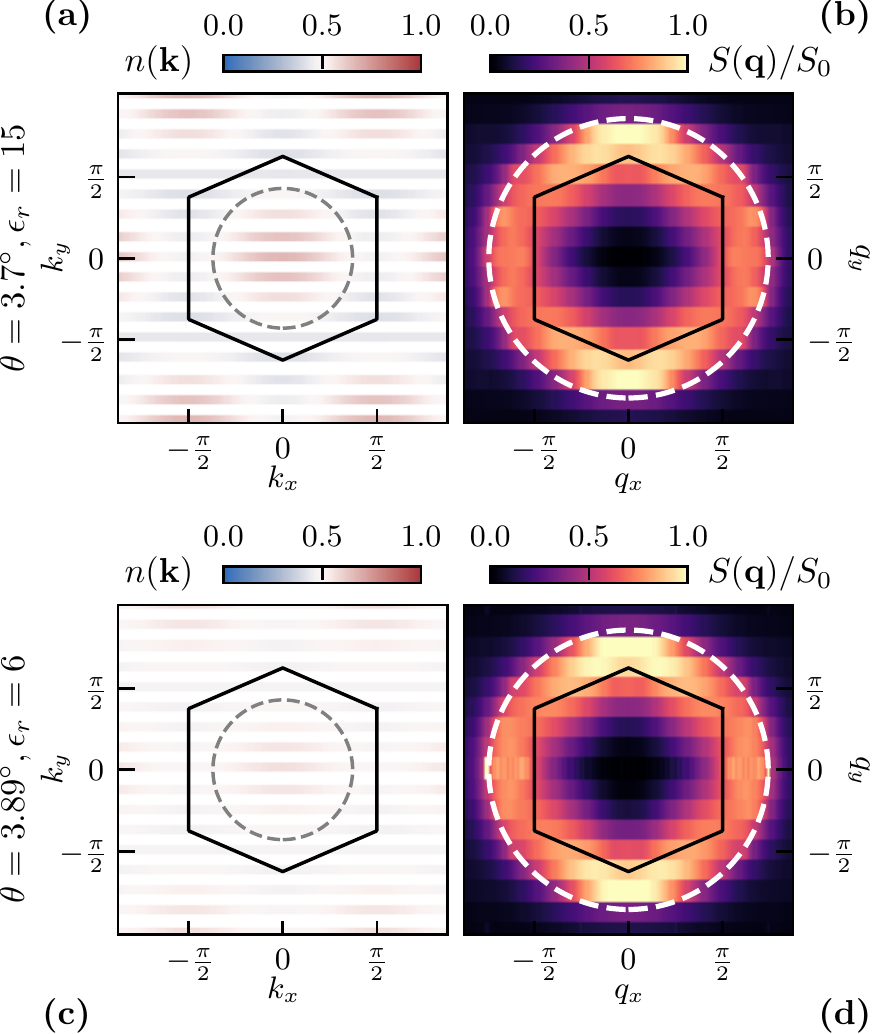}
    \caption{Comparison of DMRG electron occupations and structure factors at $L_y=8$ within the band projection approximation. The first row uses parameters $(V,\epsilon_r,\theta)=(0,15,3.7^\circ)$ and the second row uses parameters $(V,\epsilon_r,\theta)=(0,8,3.89^\circ)$. Both occupations do not contain discontinuities and nor do they change significantly with respect to bond dimension.}
    \label{fig:DMRGparamchange}
\end{figure*}

Below we will carry out numerical studies at $\epsilon_r=6$ and $8$ that are deep in the CFL phase within the band-projected Hilbert space. Within this Hilbert space, they are adiabatically connected to $(\theta,\epsilon_r) = (3.7,15)$. The fact that they belong to the same phase is confirmed in the many-body spectrum from exact diagonalization (Fig.~\ref{fig:EDparamchange}) and structure factor measurements from DMRG (Fig.~\ref{fig:DMRGparamchange}), whose interpretation will be discussed below. The parameters $(\theta,\epsilon_r) = (3.7,15)$, marked with a star in Fig. \ref{fig:HFLP}, is outside the layer-polarized phase and we expect it will have the same ground state physics in either the flat band or two band Hilbert spaces. The full phase diagram at $\nu=-1/2$ in the expanded many-band Hilbert space will be a topic for future work.

\section{Exact Diagonalization}
We perform exact diagonalization of the \textit{band projected} model Eq. (1) at different system sizes. The validity of the approximation is justified in the previous section. At half filling $\nu=-1/2$, we point out that the CFL phase can be realized in a large portion in the phase diagram $(V,\epsilon_r,\theta)$ of twisted TMD. In particular, Fig.~\ref{fig:compareLLL} lists spectra of moir\'e TMD at $(V,\epsilon_r,\theta)=(0,8,3.89^\circ)$ for various system sizes, and compares them to the spectra of Coulomb interaction in LLL of the same system sizes. The similarities between the spectra of twisted TMD and LLL strongly suggest the interacting low energy states in TMD are CFLs, because Coulomb interaction is known to stabilize the CFL phase in LLL~\cite{RezayiHaldaneCFL}. Moreover, for each system size, the momentum of low energy many-body eigenstates agree with the total momentum of compact composite Fermi seas, again indicating the phase is a CFL~\cite{scottjiehaldane}. We further confirm the CFL phase by computing the electron occupation $n(\v{k})$ in the Brillouin zone and show that it's featureless in the CFL phase but shows a Fermi surface in the FL phase. The CFL phase region is centered around $\theta \approx 3.6^\circ, V=0$ and small $\epsilon_r$, as expected due to a compromise between the $\nu = -1,0$ parent-state bandwidths. Finally, we comment on the possibility to see a CFL at $\nu=-3/4$. All the ED results performed on the LLL will be measured in arbitrary units.

\subsection{Low energy spectrum and the identification of emergent Fermi sea}
\begin{figure*}
   \centering
   \includegraphics[width = \linewidth]{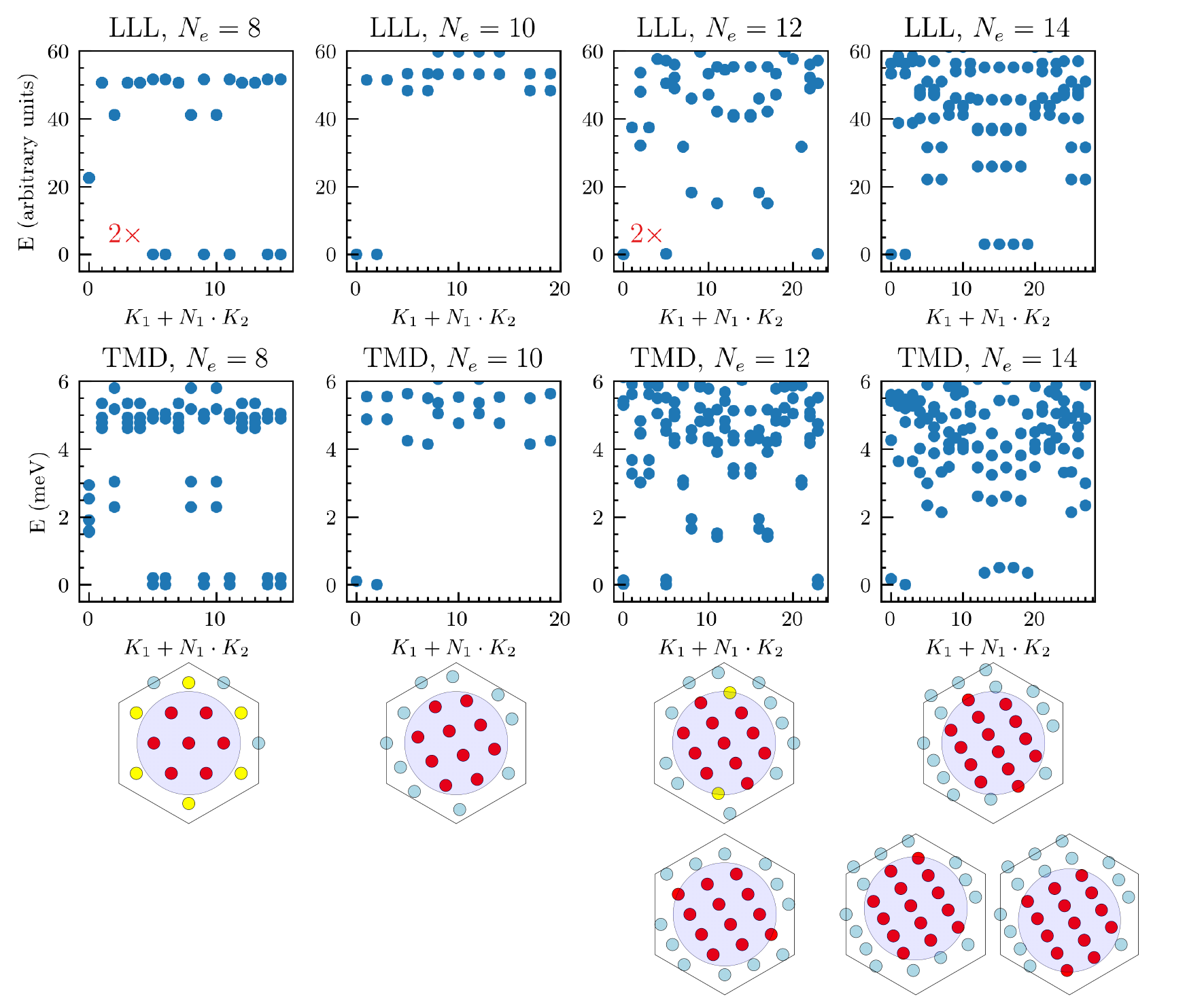}
   \caption{Comparison between the spectrum of half filled LLL with Coulomb interaction (top row) and that of twisted \ce{MoTe2} (middle row) for different system sizes at $(V,\epsilon_r,\theta)=(0,8,3.89^\circ)$. The third and fourth row illustrate the composite fermion configuration for the low energy states. We note each point in the spectrum shown in the LLL in $N_e=8$ and $N_e=12$ are two fold degenerate, related to the continuous center of mass magnetic translation symmetry; the center of mass partner differs their momentum by $(\delta K_1, \delta K_2)=(2,0)$ for $N_e=10$ and $N_e=14$.}
   \label{fig:compareLLL}
\end{figure*}

The similarities between the spectrum of LLL and moir\'e TMD, shown in Fig.~\ref{fig:compareLLL}, strongly suggest the ground states of twisted TMD are CFLs. Here we analyze the ED spectrum in more detail and explain why they point to the emergent Fermi sea.

To compare the spectrum between LLL and moir\'e TMD, we adopt the same real space geometry on these two systems. The vectors that generate the $N_1 \times N_2$ torus are $\v{L}_i=N_i\v{a}_i$, written in terms of the lattice vectors $\v{a}_1$ and $\v{a}_2$. We define the corresponding reciprocal lattice vectors as $\v{b}_1$ and $\v{b}_2$ in which $\v{a}_i\cdot\v{b}_j=2\pi\delta_{ij}$. Since we have chosen identical geometries for these two systems, we can also choose $\v{a}_{1,2}$ to denote the basis vectors for the magnetic unit cell in LLL as well as the basis vectors for unit cell in moir\'e TMDs. We start  by reviewing translation symmetries and features of the CFL spectrum in LLL; we will later find that similar discussions apply to moir\'e TMD straightforwardly. We choose the magnetic field such that there is one flux quantum per unit cell, so the number of fluxes $N_\phi=N_1 N_2$.

We first review the magnetic translation boundary conditions in LLL. Without loss of generality, we impose arbitrary twisted boundary conditions on our Hilbert space by threading fluxes $\phi_i$ through the non-contractible loops of the torus generated by $\v{L}_i$. This choice of boundary conditions then defines the single particle Hilbert space, in which every state satisfies the relation~\cite{haldanetorus1,haldanetorus2}
\begin{equation}
   T_{\v{L}_i}^{\textrm{mag}}\psi(\v r) = e^{-i \varphi_{\v{L}_i}(\v r)} \psi(\v r + \v L_i) =(-1)^{N_\phi}e^{i\phi_i}\psi( \v r).
   \label{eq:defbc}
\end{equation}
Here, as in the previous section, the magnetic twist $e^{-i \varphi_{\v L}(\v r)}$ distinguishes magnetic translations from ordinary translations.
We emphasize that making a choice of boundary conditions breaks the continuous magnetic translation symmetry of the LLL, and generally applying an arbitrary magnetic translation to each electron in the state will change the boundary condition~\cite{haldanetorus1,haldanetorus2,Jie_MonteCarlo}. We can use this property to relate states with different $\phi_i$ to each other.

We then define boundary conditions for composite fermions. To analyze composite fermion boundary conditions we must choose a trial wavefunction to begin with. We choose the widely accepted Read-Rezayi wavefunction, reproduced below~\cite{rezayiFermiliquidlikeStateHalffilled1994}:
\begin{equation}
   \Psi(\{ \v r_n \}) = \mathcal{P}_{\text{LLL}} \det_{mn}\left(e^{i\tilde{\v{k}}_m\cdot \v{r}_n}\right) \Psi_{L}(\{ \v r_n \})
   \label{eq:CFL_LLL_wavefunction}
\end{equation}
This wavefunction can be understood from a parton construction similar to the one in the main text. The first term, $\Psi_f = \det\left(e^{i\tilde{\v{k}}_m\cdot \v{r}_n}\right)$, describes a composite Fermi sea of fermions that feel no flux. The second term $\Psi_{L}(\{ \v r_n \})$ describes a bosonic Laughlin state of bosons that do feel the electronic flux. We will see that the composite fermion boundary conditions, manifesting through the quantization of the CF momenta $\tilde{\v{k}}_m$, can be completely different than the electron boundary conditions. This is because the CFs boundary condition is set by the \textit{emergent} gauge flux through the torus handles, which is \emph{dynamical}.

To impose the boundary conditions \eqref{eq:defbc}, we note that individual magnetic translations $T_{j,\v{L}_i}^{\textrm{mag}}$ commute with the projector to the LLL: $[T^{\textrm{mag}},{\mathcal P}_{LLL}]=0$. We then apply the magnetic translation on the $j$th particle in \eqref{eq:CFL_LLL_wavefunction}. By using the commutativity of magnetic translation and LLL projection we arrive at,
\begin{equation}
   T_{j,\v{L}_i}^{\textrm{mag}}\Psi = \mathcal{P}_{\text{LLL}} \left[\left(t_{j,\v{L}_i}\Psi_f\right) \left(T_{j,\v{L}_i}^{\textrm{mag}}\Psi_{L}\right)\right],
\end{equation}
where $t_{j,\v{L}_i}$ is an ordinary translation, with no magnetic twist. Here we have placed the magnetic twist factor onto the translation of the Laughlin state, anticipating the fact that the Laughlin state can be symmetric under magnetic translations while the zero-flux fermion wavefunction is symmetric under ordinary translations.

We suppose that the bosonic Laughlin wavefunctions satisfy boundary condition $\phi^{(b)}_i$, not necessarily equal to $\phi_i$:
\begin{equation}
   T_{j,\v{L}_i}^{\textrm{mag}}\Psi_{L} = (-1)^{N_\phi} e^{i\phi^{(b)}_i}\Psi_{L} \label{defmagbc}
\end{equation}
Note that Laughlin states have a two-fold degeneracy due to the part of the center of mass translation symmetry. This degeneracy directly translates to the center of mass degeneracy of the composite fermion wavefunction~\eqref{eq:CFL_LLL_wavefunction}.

We now impose boundary conditions on the composite fermion part of the wavefunction
\begin{equation}
\begin{aligned}
   t_{j,\v{L}_i}^{\textrm{mag}} \Psi_f 
   & = \left(\sum_{P}(-1)^P e^{i\tilde{\v{k}}_{P(j)}\cdot\v{L}_i}e^{i\tilde{\v{k}}_{P(m)}\cdot (\v{r}_n)}\right) \\
   & = e^{i\phi^{(f)}_{ i}}\mathcal{P}_{\text{LLL}} \det_{mn}\left(e^{i\tilde{\v{k}}_m\cdot \v{r}_n}\right) ,
\end{aligned}
\end{equation}
which corresponds to $e^{i \v k_j \cdot \v L_i} = e^{i \phi^{(f)}_{i}}$. The quantization of composite fermion momentum is thus
\begin{equation}
   \tilde{\v{k}}=\frac{n_1 +\frac{\phi_1^{(f)}}{2\pi}}{N_1} \v{b}_1+\frac{n_2 +\frac{\phi_2^{(f)}}{2\pi}}{N_2} \v{b}_2.\label{defcfk}
\end{equation}

The boundary conditions of the two sub-wavefunctions then sum to the electron boundary conditions
\begin{equation}
   \phi_i = \phi_i^{(f)} + \phi_i^{(b)} \label{factorizationbc}.
\end{equation}

Indeed, the boundary condition for electrons --- corresponding to the electromagnetic fluxes through the handles of the torus --- are split up between the CFs and the composite bosons that form the Laughlin state. The ``relative" degree of freedom, corresponding to increasing $\phi_i^{(f)}$ while decreasing $\phi_i^{(b)}$, such that $\phi_i$ is held fixed, corresponds to the dynamical flux of the emergent gauge field. 

We now compute the center of mass momentum for the state $\eqref{eq:CFL_LLL_wavefunction}$. To do this, we define the center of mass momentum $\v{K}$ for a state as
\begin{equation}
   \left(\prod_{j} T_{j,\v{a}_i}^{\textrm{mag}}\right) \Psi = e^{i \v{K}\cdot \v{a}_i} \Psi. \label{defcomM}
\end{equation}
This definition is such that there always exists a bosonic Laughlin state which, when periodic boundary condition $\phi_i^{(b)}=0$ is imposed, has zero center of mass momentum $K_1=K_2=0$.

Starting with this canonical definition of COM momentum at periodic boundary condition, the COM momentum at arbitrary boundary condition can be obtained from flux insertion:

\begin{equation}
   \v{K}(\phi) = N_e\left( \frac{\phi_1}{2\pi N_1} \v{b}_1+\frac{\phi_2}{2\pi N_2}\v{b}_2\right) + \v{K}(\phi=0).\label{KphiK0}
\end{equation}

Following \eqref{defcomM}, we are ready to compute the center of mass momentum for the state $\eqref{eq:CFL_LLL_wavefunction}$:
\begin{equation}
   \begin{aligned}
       &\left(\prod_{j} T_{j,\v{a}_i}^{\textrm{mag}}\right) \Psi(\{ \v r_n \})\\
       & = e^{i\sum_m \tilde{\v{k}}_m\cdot\v{a}_i}\mathcal{P}_{\text{LLL}} \det_{mn}\left(e^{i\tilde{\v{k}}_m\cdot \v{r}_n}\right) \left(\prod_{j} T_{j,\v{a}_i}^{\textrm{mag}}\right)\Psi_{L}(\{ \v r_n \})\\
       & = e^{i(\sum_m \tilde{\v{k}}_m+\v{K}_L(\phi^{b}))\cdot\v{a}_i}\Psi(\{ \v r_n \}),
   \end{aligned}
\end{equation}
from which the COM momentum can be read out as
\begin{equation}
\begin{aligned}
   \v{K}_{\Psi} &= \sum_m \tilde{\v{k}}_m+\v{K}_L(\phi^{b}),\\
   &= \sum_m \tilde{\v{k}}_m+ \frac{N_e\phi^{(b)}_1}{2\pi N_1} \v{b}_1+\frac{N_e\phi^{(b)}_2}{2\pi N_2}\v{b}_2,
\end{aligned}
\label{eq:CFSmomentum}
\end{equation}
where we have applied \eqref{KphiK0} for the bosonic Laughlin state in the second line.

From \eqref{eq:CFSmomentum}, \eqref{defcfk} and \eqref{factorizationbc}, we notice that the COM momentum of the CFL state $\v{K}_{\Psi}$ is entirely determined by the physical boundary condition $\phi_{1,2}$ and the integer part of composite fermion momenta $\{\tilde{k}_i\}$. This means that the COM momentum is determined by $\phi_{1,2}$ and the relative configuration of the composite fermion momenta. Thus, for the purpose of interpreting CFL many body momentum from ED numerics, the concrete value of $\phi_i^{(f)}$ does not matter.

However, there could be a preferred $\phi_i^{(f)}$ energetically. We assume that $\phi_i^{(f)}$ is self-consistently chosen for each state such that the compact composite Fermi sea is centered at $\v{\Gamma}$ to minimize the variational energy, which we conjecture to be captured by a dispersion $E(\tilde{\v{k}})$ that grows with $|\tilde{\v{k}}|$ for the particle-hole symmetric CFL state in LLL. When we plot the composite Fermi sea in Fig.~\ref{fig:compareLLL}, we pick particular choices of $\phi^{(f)}_{1,2}$ for each system size such that the composite Fermi sea is centered at $\v{\Gamma}$.

We now describe the CFL COM momentum counting for all the low energy states in all system sizes. For all system sizes $N_1, N_2$ we study here, we can choose the physical boundary condition to be $\phi_i=0$.

We start with $N_e=8$ electrons on a $4\times 4$ torus. In this case, $\phi_i^{(b)}=-\phi_i^{(f)} = 0$ already gives a composite fermion Fermi sea centered at $\bm\Gamma$ point. For both LLL and twisted TMD, there are 12 low energy states living in 6 momentum sectors consisting of the ground state manifold. These twelve states, six modulo center of mass degeneracy, correspond to the six possible momenta for the eighth, highest energy, composite fermion shown in yellow in the third row of Fig. \ref{fig:compareLLL}.

We proceed and focus on $N_e=10$ electrons on a $4\times 5$ torus. We choose $\phi_2^{(b)}=-\phi_2^{(f)} = \pi$ to give a presumably low energy composite fermion configuration centered at $\bm\Gamma$. From the configuration shown in the third row $\sum \tilde{\v{k}}_i=0$, $\v{K}=N_e\phi_2^{(b)}/2\pi N_2\v{b}_2=\v{b}_2$, which corresponds to the state appearing in sector $0$. The state in sector $2$ is related to this state by center of mass translation. We numerically study the particle hole symmetry of the CFL in twisted \ce{MoTe2} using the $4\times 5$ system by calculating $|\braket{\psi_0|{\cal PH}|\psi_2}|^2=0.89$, in which $\ket{\psi_i}$ refers to the ground state in momentum sector $i$. The action of particle hole is $({\cal PH})c^\dagger_{\v{k}}({\cal PH})^{-1}=c_{\v{k}}, ({\cal PH})i({\cal PH})^{-1}=-i$. Thus the numerical particle hole breaking per particle is on the order of one percent. In contrast, $|\braket{\psi_0|{\cal PH}|\psi_2}|^2=1$ in the LLL, which is known to have perfect particle hole symmetry.

We then study $N_e=12$ electrons on a $4\times 6$ torus. We argue that two different boundary conditions must be chosen for the low energy states: $\phi_i^{(b)}=-\phi_i^{(f)} = 0 $ for the states appearing in energy sectors $5$ and $23$, shown in the third row, and $\phi_2^{(b)}=-\phi_2^{(f)} = \pi$ which appears in sector $0$, shown in the fourth row. We numerically observe these two ED states are close in energy.

We conclude by analyzing $N_e=14$ electrons on a $4\times 7$ torus. We choose $\phi_2^{(b)}=-\phi_2^{(f)} = \pi$. The symmetric Fermi sea configuration shown in the third row corresponds to the states in sector $0$, which is again related to the state in sector $2$ by the center of mass magnetic translations. On the other hand, the other low energy states, corresponding to less symmetric composite fermion Fermi seas, are shown in the fourth row.

\subsection{Explicitly LLL or ideal band projected CFL wavefunction}
Here we comment that there exists an explicitly LLL projected CFL wavefunction defined on the torus geometry~\cite{shaoprl,scottjiehaldane,Jie_MonteCarlo,Jie_Dirac}, and such a wavefunction can be easily generalized to be an ansatz for the CFL state which is explicitly projected to ideal flatbands. In LLL, such projected wavefunction reads:
\begin{eqnarray}
   \Psi_{\rm CFL} &=& \det M_{ij}\prod_{k=1}^{2} f(\sum_iz_i-\alpha_k).\label{LLLCFL}\\
   M_{ij} &=& e^{\frac{1}{8}(z_id^*_j-z_i^*d_j)}\prod_{k\neq i}^{N_e} f(z_i-z_k-d_j+\bar d),\nonumber\\
   d_i &=& \frac{mL_1+nL_2}{N_e} + \frac{L^{(f)}}{N_e};\quad\sum_{k=1}^{2}\alpha_k = L^{(b)},\nonumber
\end{eqnarray}
where $d_i$ is the complex coordinate of the composite fermion dipole $d_i = d_i^x + id_i^y$, and $z=x+iy$ is the complex coordinate of electron. In LLL, the dipole is related to the momentum of composite fermion via dipole-momentum locking $d^a=2l_B^2\epsilon^{ab}\tilde k_b$. Since the reciprocal lattice is related by a $90^\circ$ rotation and appropriate rescaling, $\bm b_{i}\rightarrow \epsilon_{ij}\bm a_j$, the dipole is quantized on a $L_{1,2}/N_e = 2 L_{1,2}/(N_1 N_2)$ lattice and shifted by an internal flux $L^{(b/f)}=\left[-\phi^{(b/f)}_2L_1+\phi^{(b/f)}_1L_2\right]/2\pi$. The $\bar d$ is a parameter of the wavefunction and can be chosen as center of dipoles $\bar d=\sum_id_i/N$. In the above $\alpha_{1,2}$ labels the COM zeros that determines the COM momentum of the bosonic Laughlin state, which is shifted by the Laughlin state's boundary condition. To obtain a torus wavefunction, we need to use Weierstrass sigma functions $\sigma(z)$ that is quasi-periodic in $L_{1,2}$, and we have defined $f(z)=\sigma(z)\exp(-|z|^2/4N_\phi)$~\cite{haldaneholomorphic,haldanemodularinv,Jie_MonteCarlo}.

The link between the Read-Rezayi wavefunction \eqref{eq:CFL_LLL_wavefunction} and the projected wavefunction \eqref{LLLCFL} can be understood as follows. First, the Jastrow factor from the bosonic Laughlin wavefunction can be pulled into the determinant, so we can rewrite $\Psi_{f}\Psi_{L}$ as follows~\cite{JainKamilla},
\begin{equation}
   \left(\det_{ij}e^{i\bm k_j\cdot\bm r_i}\right)\prod_{i<j}\left(z_i-z_j\right)^2 = \det_{ij}\left[e^{i\bm k_j\cdot\bm r_i}\prod_{k\neq i}\left(z_i-z_k\right)\right].\nonumber
\end{equation}
Notice the Bloch factor $e^{i\bm k\cdot\bm r} = e^{i\bm d\times\bm r/2} = e^{(\bar dz-d\bar z)/4}$. We then approximate the LLL projection by replacing the anti-holomorphic coordinate $\bar z$ with the holomorphic partial derivative $l_B^2 \partial_z$ such that $P_{\rm LLL}\left(\Psi_f\Psi_L\right)$ is approximated by $\det_{ij} \left(e^{\frac14 \bar d_j z_i} \prod_{k\neq i}  (z_i-z_k-d_j)\right)$. Finally translating into torus geometry arrives at \eqref{LLLCFL}. See Ref.~\cite{scottjiehaldane} for details. Generalizing this wavefunction to $1/2m$ filling is straightforward~\cite{Jie_Dirac}.

In the end we comment that the exact mapping between LLL and ideal band mentioned in the main text, Eq. (3), enables an explicit construction of CFL variational wavefunction without magnetic field: $\Psi_{\rm CFL}^{\rm band} = \Psi_{\rm CFL}(\{z_i\})\prod_i e^{-K(\v r_i)} \zeta_{l_i}(\v r_i)$. We leave a detailed study of this flat band CFL wavefunction for the future.

\subsection{Phase Diagrams and Phase Identification}
\begin{figure*}
   \centering
   \includegraphics[width = \linewidth]{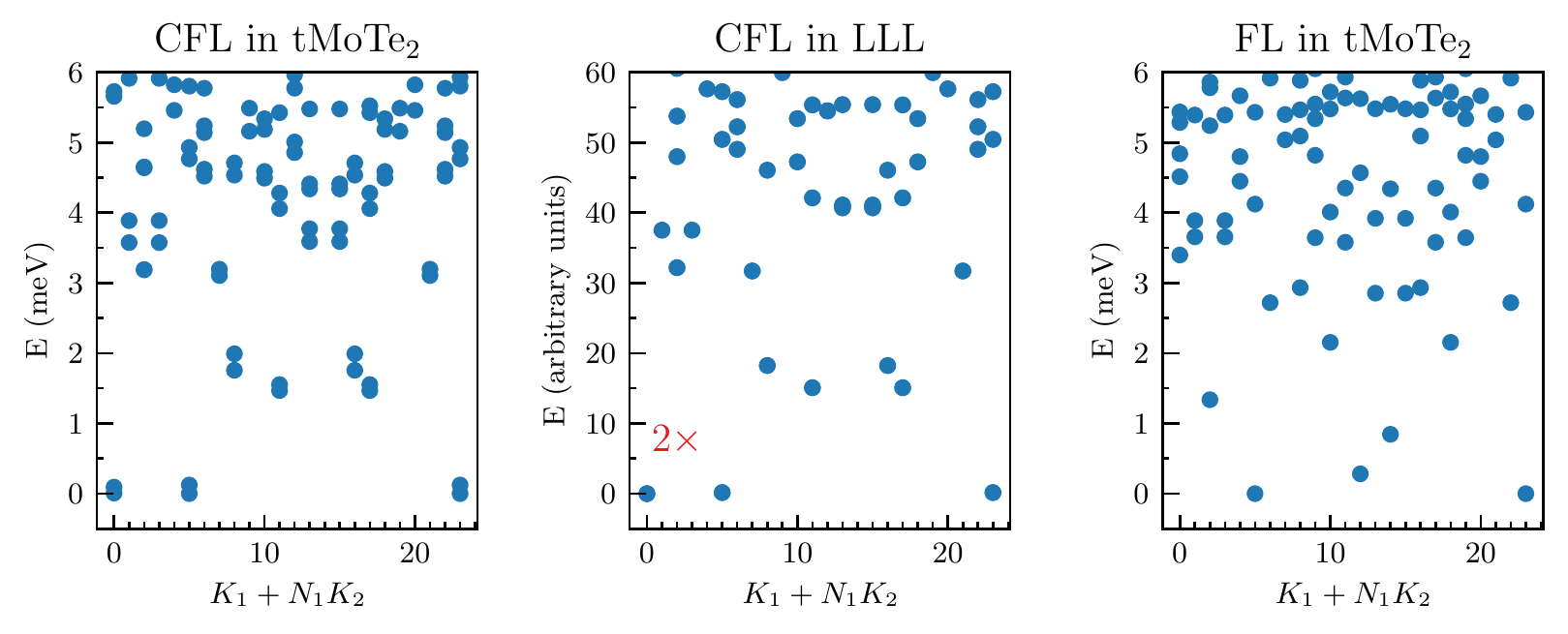}
   \caption{Representative energy spectrum for the $\nu=-1/2$ CFL phase in twisted \ce{MoTe2}, the CFL phase in LLL, and the FL phase in twisted \ce{MoTe2}.
   All the ED are performed on $4\times 6$ torus with the same geometry. Notably, for CFL phase the two-fold COM degeneracy is exact in LLL, whereas it is weakly broken in twisted MoTe$_2$. When the system transitions into the FL phase, the COM degeneracy is no longer present. 
   }
   \label{fig:transitionFL}
\end{figure*}

\begin{figure*}
   \centering
   \includegraphics[width = \linewidth]{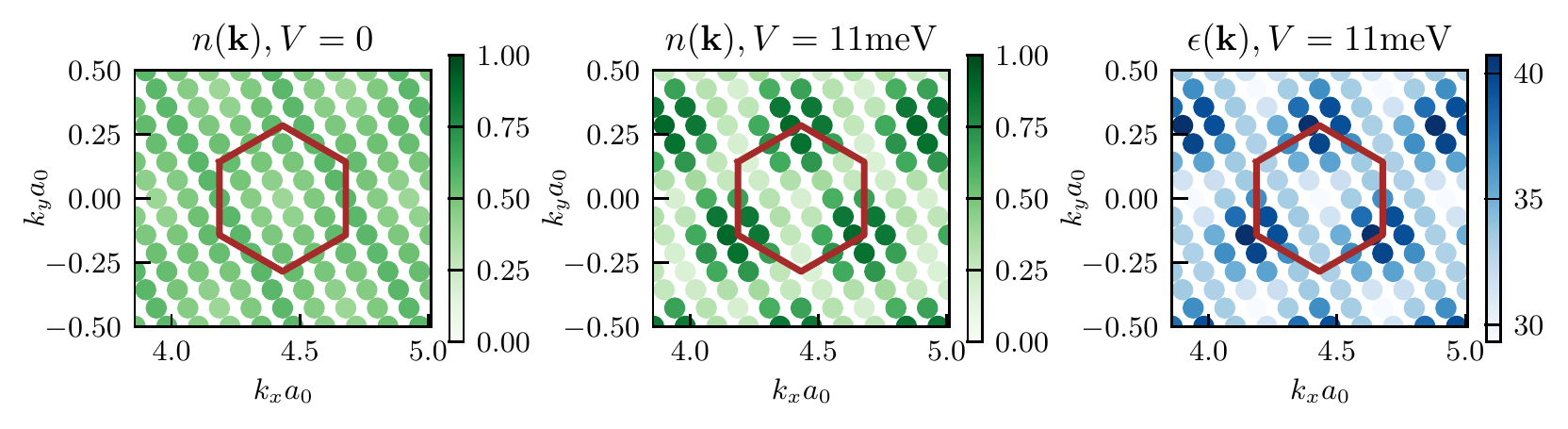}
   \caption{Occupations of holes in the ED ground state in the CFL phase and the FL phase by tuning displacement field, respectively. All the ED are performed on $4\time 6$ torus with the same geometry. The FL occupations match excellently to the Fermi distribution $\theta(\epsilon-\mu)$ for holes. }
   \label{fig:occupations}
\end{figure*}

In this section, we discuss how to identify different phases from ED. We use multiple indicators, including:
\begin{itemize}
   \item Momentum of low energy states.

   In the previous section, we have discussed the momentum of low energy states from the CFL phase. They correspond precisely to the total momentum of composite fermions Fermi sea. Moreover, low energy states of CFL are nearly degenerate, corresponding to the exact COM degeneracy in LLL. For instance, for $N_e=12$ particles, the low energy states of CFL has $K=K_1+N_1K_2=0,5,23$ and each of them are nearly two fold degenerate. Their composite Fermi sea configuration is plotted in Fig.~\ref{fig:compareLLL}.

   Differently, Fermi liquids do not exhibit any signatures of COM degeneracy. See Fig.~\ref{fig:transitionFL} for illustrations.
    
   \item The occupation number of ground state $n(\bm k)$.

   We can also identify the FL phase by computing the ground state occupation of momentum eigenstates in the Brillouin zone. In the CFL phase, the ground state occupations are relatively uniform throughout the Brillouin zone and there are no significant Fermi surface discontinuities in the occupation. On the other hand, in the FL phase, the ground state occupations not only exhibit a discontinuous feature at the putative Fermi surface, but also closely follow the non-interacting Fermi distribution $n_{(\epsilon,\mu)}=\theta(\epsilon-\mu)$ for holes very well, as shown in Fig.~\ref{fig:occupations}(b) and (c). The quasiparticle residue $Z$, which characterizes the discontinuity at the Fermi surface, is not $1$ as the noninteracting FL because it is renormalized by the interactions. We note that this phase identifier is further supported by iDMRG results shown in Fig. 2(a), in which the CFL clearly shows a featureless Brillouin zone in terms of electron occupations, whereas the FL has a clear Fermi surface.
\end{itemize}

\begin{figure*}
   \centering
   \includegraphics[width = 0.7\linewidth]{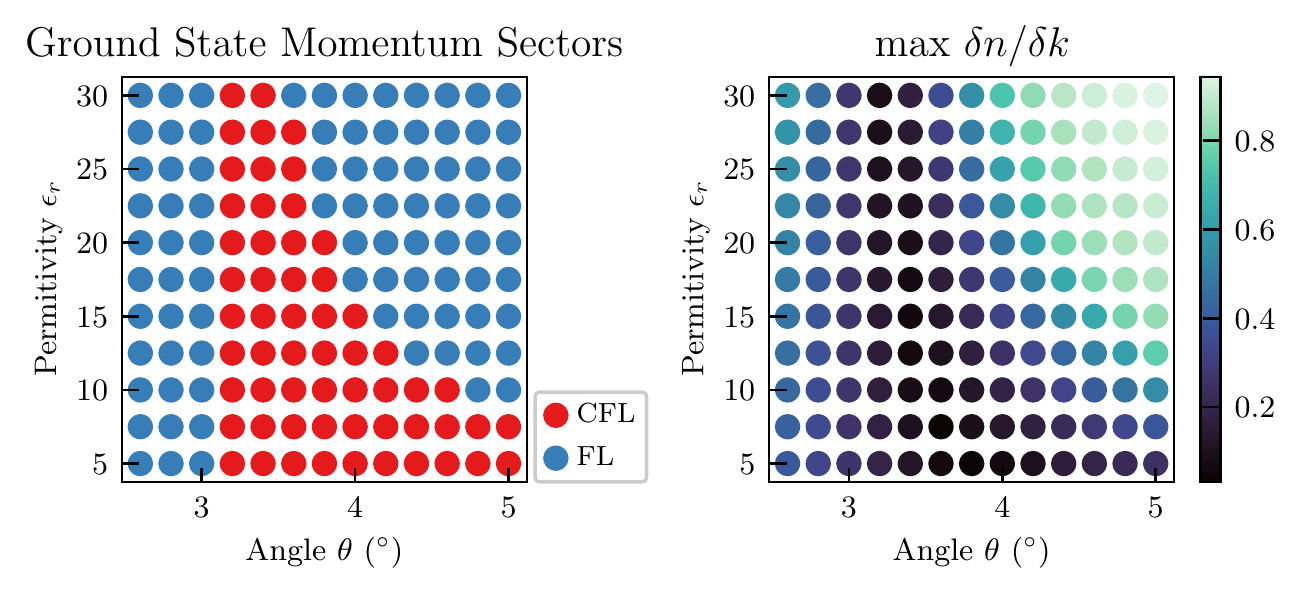}
   \caption{Band-projected phase diagram of interacting twisted \ce{MoTe2} at $\nu=-1/2$ at zero displacement field obtained by two different indicators from ED on a $4\times 6$ torus. The phase diagram of left figure is resolved by the momentum structure of low energy states while that on the right is resolved from the occupation $\braket{n({\bm k})}$. See text for details.}
   \label{fig:phase_diagram_comparison}
\end{figure*}

We cross-check the two methods by computing the zero displacement field phase diagram in the $(\theta,\epsilon_r)$ plane, shown in Fig.~\ref{fig:phase_diagram_comparison}. In this phase diagram, only the CFL and FL phases appear. The phase diagram computed by the low energy momentum sectors is shown in Fig.~\ref{fig:phase_diagram_comparison}(a), in which the CFL phase is shown in red dots and the FL phase is shown in blue dots. The phase diagram computed by the finite differences in occupation is shown in Fig.~\ref{fig:phase_diagram_comparison}(b). The dark regions represent a small variation in occupations and thus represent the CFL phase. The bright regions represent the FL phase. The shapes of the phase boundaries agree with each other. The large volume of the CFL state in this band projected phase diagram implies that our result is robust to parameter changes both in twist angle variations and strength of interaction, supporting the possibility that the CFL state can be realized in realistic experiments.

In Fig. 3(b) we identify the phases in the $(\theta,V)$ plane using the many-body momentum sectors. We keep $\epsilon_r=8$ throughout. There is clearly a large portion of the phase diagram exhibiting CFL-like features, represented by the red region centered around $3.6^\circ$ with a small displacement field. The displacement field drives a phase transition from the CFL into the FL phase, shown in green.

We mark the topological transition line in gray and remark that near the topological phase transition the band gap closes so the band projection approximation in our ED procedure no longer works. We emphasize that this phase diagram is obtained assuming band projection throughout the parameters used in the phase diagram.

\subsection{$\nu=-3/4$ Composite Fermi Liquid and Associated Phase Diagrams}
\begin{figure*}
   \centering
   \includegraphics[width = \linewidth]{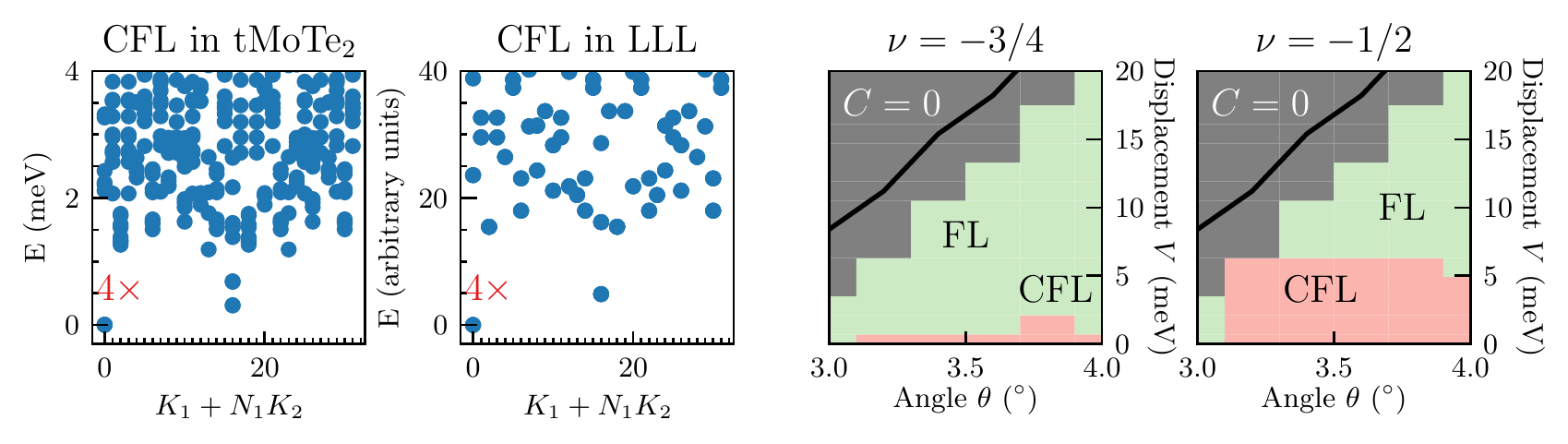}
   \caption{The first two panels compare the spectrum of twisted \ce{MoTe2} and the LLL at $\nu=-3/4$ within the band-projected approximation. The four fold center of mass degeneracy is broken weakly in twisted \ce{MoTe2}, but is too close to be resolved by eye in the ground state. We then compare the band-projected phase diagrams at $\epsilon_r=8$ and $\nu=-3/4$ and $\nu=-1/2$. The large extent of the composite Fermi liquid phase is shown in red. The green regions correspond to Fermi liquid states, and the gray regions are near or past the $C=1\to0$ topological transition marked with a black line. The composite Fermi liquid state is much more unstable to displacement field at $\nu=-3/4$ compared to $\nu=-1/2$.}
   \label{fig:4x8}
\end{figure*}

\begin{figure*}
   \centering
   \includegraphics[width = 0.7\linewidth]{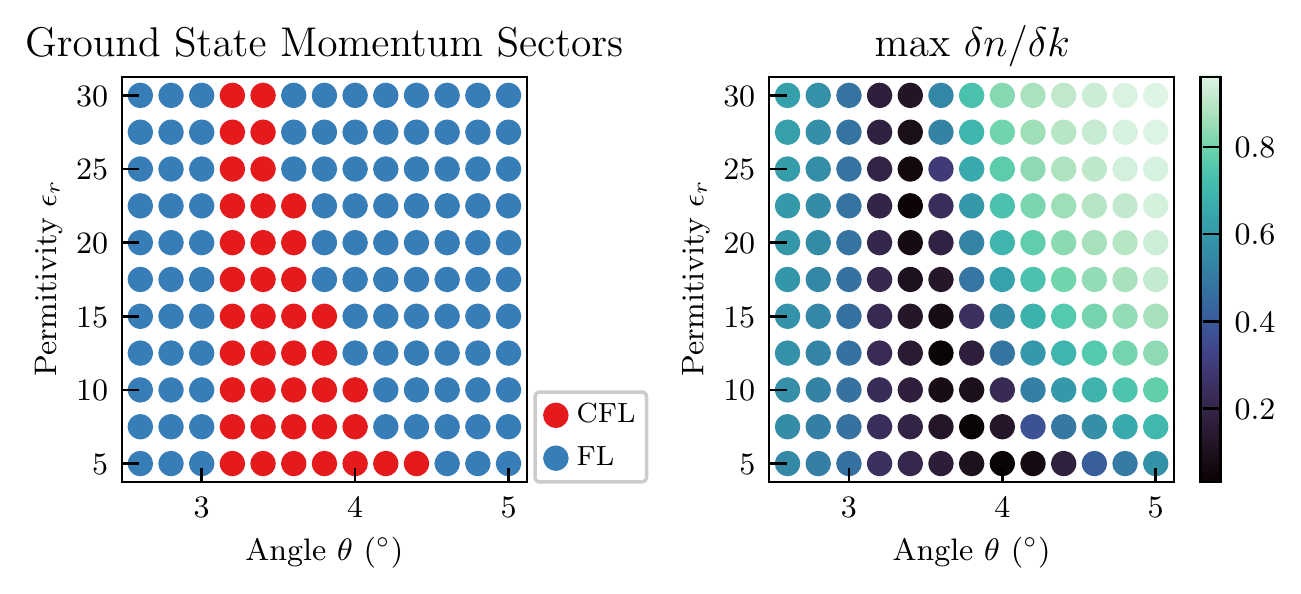}
   \caption{Band-projected phase diagram of interacting twisted \ce{MoTe2} at $\nu=-3/4$ at zero displacement field obtained from two different indicators from ED on a $4\times 8$ torus. The phase diagram of left figure is resolved by the momentum structure of low energy states while that on the right is resolved from the occupation $\braket{n({\bm k})}$. See text for details.}
   \label{fig:4x8_phase_diagram_comparison}
\end{figure*}
The possibility of a $\nu=-3/4$ CFL is investigated also using ED on a $4\times 8$ torus. Since such a state is closer to $\nu=-1$, we expect that it generally has a larger spin gap than the possible $\nu=-1/2$ CFL. This means that the valley polarization we assume in the ED procedure is better justified at this filling. However, the layer-polarization instability discussed above will modify this phase diagram in the larger Hilbert space of multiple valence bands.

We compare the spectrum between the $\nu=-3/4$ CFL and the $\nu=3/4$ CFL from the LLL using dual gate screened Coulomb interactions in Fig.~\ref{fig:4x8}. Each point in the LLL spectrum has a four fold center of mass degeneracy, which is weakly broken in twisted \ce{MoTe2}. The similarity of the spectrum suggests that a CFL where $8\pi$ fluxes are attached to each electron can also be formed at $\nu=-3/4$.

In the last two panels of Fig.~\ref{fig:4x8} we show the many-body phase diagram at $\nu=-3/4$ and at $\nu=-1/2$ assuming band projection. Since $\nu=-3/4$ is closer to $\nu=-1$ we conjecture that the valley polarization will be more stable to the displacement field. We note that the $\nu=-3/4$ CFL region (red) has significantly shrunk compared to the $\nu=-1/2$ CFL, and most of the phase diagram of the $C=1$ band is a FL phase (green). We mark the topological transition within the top valence band by a  black  line, and remark that near the topological phase transition the band gap closes so the band projection approximation in our ED procedure is no longer necessarily (gray region). The layer polarization instability from the second band may occur significantly below this line when a larger Hilbert space is considered.

\section{Density Matrix Renormalization Group}

This Appendix discusses the infinite density matrix renormalization group (iDMRG) numerics used in the main text. We first discuss the computational basis used, then express the structure factor in this basis, and finally give additional details on the phase identification including finite size scaling.

\subsection{Computational Basis}
For DMRG we consider a moir\'e lattice geometry 
\begin{align}
 \v{a}_1 &= a_M \left( \frac{\sqrt{3}}{2}, \frac{1}{2} \right), \quad
 \v{a}_2 = a_M \left( 0, 1 \right)\\
 \v{g}_1 &= \frac{2\pi}{a_M} \left( \frac{2}{\sqrt{3}}, 0 \right), \quad
 \v{g}_2 = \frac{2\pi}{a_M} \left( -\frac{1}{\sqrt{3}}, 1 \right).
\end{align}
where $a_M = \n{\v{a}_1}$ is the moir\'e length. We consider rectangular Brillouin zone $\mathcal{B} = [-G_x/2,G_x/2] \times [-G_y/2 \times G_y /2]$ where $G_x = \frac{4\pi}{\sqrt{3}a}, G_y = \frac{2\pi}{a}$ so that $\v{a}_1 = 2\pi\left(  \frac{1}{G_x}, \frac{1}{2G_y} \right)$. 

We solve the kinetic Hamiltonian Eq. \eqref{eq:hole_bandstructure_app} as
\begin{equation}
   h \ket{\phi^{\v{k} a}} = \epsilon^{\v{k} a} \ket{\phi^{\v{k} a}}, 
\end{equation}
where $\ell = {B,T} = {-1,+1}$ labels layers, and $a$ labels bands. We impose boundary conditions in the $y$ direction corresponding to the momenta discretization
\begin{equation}
 k_y[n] = G_y \begin{cases}
   -\frac{1}{2} + \frac{n + \frac{\Phi_y}{2\pi}}{L_x} & \text{if } L_y \equiv 0 \pmod 2\\
   -\frac{1}{2} + \frac{n+ \frac{\Phi_y}{2\pi}+\frac{1}{2}}{L_x} & \text{if }  L_y \equiv 1 \pmod 2 \\
 \end{cases}.
\label{eq:DMRG_wire_momenta}
\end{equation}
We therefore have $L_y$ evenly-spaced``wires" across the Brillouin zone. The value of $L_x \gg 1$ does not affect the physics, provided that it is sufficiently large. If $\Phi_y = 0$ then one wire will go through $\Gamma$, giving electrons periodic boundary conditions (PBC); $\Phi_y = 1/2$ corresponds to anti-periodic boundary conditions (APBC).

Following previous DMRG work on moir\'e systems~\cite{MPOCompression2, parker2021strain, Dan_parker21, wang2022kekul}, we take a computational basis of hybrid Wannier functions that are maximally (exponentially) localized along $\v{a}_1$ and periodic along $\v{a}_2 \propto \hat{y}$. Let $\hat{P}$ be a projector to the band (or bands) of interest. We transform to the Wannier-Qi states~\cite{WannierQi}
\begin{equation}
  \ket{w^{n,k_y,b}} = \frac{1}{\sqrt{L_x}} \sum_{k_x} \ket{\varphi^{\v{k} b}} e^{-i \v{k} \cdot \v{a}_1 n}, \ket{\varphi^{\v{k} b}}  =     \ket{\phi^{\v{k} a}} U^{ab}(\v{k})
\end{equation}
where $U_{nb}(\v{k})$ is a unitary matrix that transforms into a Wannier gauge such that
\begin{subequations}
   \begin{align}
       \hat{T}_{1} \ket{w^{n,k_y,b}} &= \ket{w^{n-1,k_y,b}}\\
       \hat{T}_{2} \ket{w^{n,k_y,b}} &= e^{i 2 \pi k_y/G_y} \ket{w^{n,k_y,b}}\\
       \hat{P} e^{-i \hat{x} G_x} \hat{P} \ket{w^{n,k_y,b}} &= e^{i 2\pi [n+P_x(k_y,b)]} \ket{w^{n,k_y,b}}
   \end{align}
\end{subequations}
where the polarizations $P_x(k_y,b) = \frac{G_x}{2\pi} \braket{w^{0,k_y,b}|\hat{x}|w^{0,k_y,b}}$ are the centers of the Wannier orbitals in the first unit cell, in accordance with the modern theory of polarization~\cite{vanderbilt2018berry}. The polarizations for the top two valence bands of the $K$-valley are shown in Fig. \ref{fig:app_bandstructure_quantum_geometry}(b). These have winding $C = \int dk_y \frac{dP_x}{dky} = +1,-1$ for the respectively, matching the Chern number~\cite{vanderbilt2018berry}. We now specialize to the case where $\hat{P}$ projects to the top $C=+1$ valence band of the $K$-valley (though this method applies equally well to both valleys and multiple bands, which will be the subject of future work).

At the many-body level, we consider Fermi creation operators
\begin{equation}
   \ket{\varphi^{\v{k}}} = \hat{c}^\dagger_{\v{k}} \ket{0},\quad  \hat{c}^\dagger_{\v{k}}
= \frac{1}{\sqrt{L_x}} \sum_{n} e^{i \v{k} \cdot \v{a}_1 n} \hat{d}^\dagger_{n,k_y},
\end{equation}
which correspond to Bloch states and Wannier-Qi states in the top valence valence band of the $K$ valley, respectively. We then express the Hamiltonian Eq. \eqref{eq:app_TMD_hamiltonian} in the $\hat{d}^\dagger_{n,k_y}$ basis, which should be interpreted as the putting the TMD Hamiltonian onto a cylinder geometry that is infinite along the $x$ direction and periodic along the $y$ direction with circumference $L_y$. The choice of maximally-localized Wannier orbitals along $x$ ensures that hoppings decay as fast as possible in the $x$-direction. However, the quasi-long range screened Coulomb interactions decay quite slowly along $x$.

To perform infinite DMRG, one must express the Hamiltonian as a matrix product operator (see e.g.~\cite{pirvu2010matrix}), which is a tensor network composed of size $2\times 2 \times D \times D$ tensors where $2$ is the physical dimension (for fermions) and $D$ is the virtual ``bond dimension''. One may naively construct an MPO from a Hamiltonian in the $d^\dagger_{n,k_y}$ basis by imposing a finite cutoff $M > 0$ on the interaction range along the cylinder. However, this has two serious drawbacks: (i) the MPO bond dimension increases as $O(M^2)$, reaching $D \sim 2000$ by $M = 2$ at $L_y =6$~\cite{MPOCompression2} and (ii) the relative stability of various many-body ground states (particular quantum Hall states) are known to be sharply modified by the interaction distance. To circumvent this issue, some of us developed a technique of ``MPO Compression''~\cite{MPOCompression1,MPOCompression2}, which creates a faithful representation of the MPO, including its quasi-long range interactions. In practice we ensure that the error of each matrix element of the Hamiltonian is $< 10^{-3}$ \si{meV} for up to 200 sites along the 1D chain. This technique has been previously applied to twisted bilayer graphene in the cases of 1, 2, 4, or 8 degrees of freedom per $k$-point~\cite{MPOCompression2,parker2021strain,Dan_parker21,wang2022kekul}. In this single-DOF case, the resulting bond dimensions are $D\sim 130 - 250$ for $L_y = 5-10$ respectively.

\subsection{Structure Factor}

To identify a composite Fermi liquid, we compute the structure factor, i.e. the density-density correlation function. The standard structure factor is defined as
\begin{eqnarray}
   S_0(\v{q}) = \braket{\hat{\rho}_{\v{q}} \hat{\rho}_{-\v{q}}}
   \label{eq:structure_factor}
\end{eqnarray}
in terms of the density operator \eqref{eq:app_density_operator}. However, the dominant features of $S(\v{q})$ here --- and in any lattice system --- are peaks at the Bragg lattice from $\braket{\hat{\rho}_{\v{q}}} = \sum_{\v{G}} \delta_{\v{G}\v{q}} \braket{\hat{\rho}_{\v{G}}}$, where $\v{G}$ is a reciprocal lattice vector. To identify the non-trivial features associated to a (composite) Fermi surface, we focus on the connected part of the structure factor
\begin{equation}
   S(\v{q}) = \braket{\hat{\rho}_{\v{q}} \hat{\rho}_{-\v{q}}} - \braket{\hat{\rho}_{\v{q}}}\braket{\hat{\rho}_{-\v{q}}}.
   \label{eq:connected_structure_factor}
\end{equation}
We now resolve this in the computational basis $\hat{d}_{n,k_y}$. A straightforward, albeit lengthy, computatation shows
\begin{widetext}
\begin{subequations}
\begin{align}
   S(\v{q})
   &= \sum_{\substack{n_0 \in \mathbb{Z}\\ \n{\Delta N} \le M\\ \n{\Delta n}, \n{\Delta n'} \le N}}
   \sum_{k_y,k_y'} e^{i\v{q} \cdot \v{a}_1 (\Delta N + \Delta n' -\Delta n)}
   \tilde{\Lambda}_{\v{q}}(\Delta n, k_y), \tilde{\Lambda}_{-\v{q}}(\Delta n', k_y')
   M_{\Delta n, k_y, \Delta n', k_y', \Delta N, q_y}\\
   \tilde{\Lambda}_{\v{q}}(\Delta n, k_y) &= \sum_{k_x} e^{-i\v{k}\cdot \v{a}_1 \Delta n} \Lambda_{\v{q}}(k_x,k_y)\\
   M_{\Delta n, k_y, \Delta n', k_y', \Delta N, q_y}
   &= 
   \braket{
   \hat{d}_{n_0, k_y}^\dagger \hat{d}_{n_0+\Delta n,[k_y+q_y]}
   \hat{d}_{n_0+\Delta N, k_y'}^\dagger \hat{d}_{n_0+\Delta N +\Delta n',[k_y-q_y]}
   }\\
   &\hspace{1cm} -
   \braket{
   \hat{d}_{n_0, k_y}^\dagger \hat{d}_{n_0+\Delta n,[k_y+q_y]}}
   \braket{\hat{d}_{n_0+\Delta N, k_y'}^\dagger \hat{d}_{n_0+\Delta N +\Delta n',[k_y-q_y]}}.\nonumber
\end{align}
\end{subequations}
\end{widetext}
Here $[k_y+q_y]$ denotes that the quantity lies within the first Brillouin zone, and $M >0$, $N>0$ are finite-size cutoffs. As it is computationally expensive to compute $L_y^3 (2M+1) (2N+1)^2$ observables in DMRG, we restrict to $N = 1$ (which is sufficient for small $\n{q_y} \lesssim G_y$), and take several hundred values of $\Delta N$, $q_x$, which is sufficient to resolve fine features in Fourier space. 

\subsection{DMRG Phase Identification}

This section details the phase identification of the Fermi liquid and composite Fermi liquid in DMRG. We focus on system sizes $L_y = 5, 8, 10$ and bond dimensions $\chi = 1024 - 4096$, which are sufficiently large to determine the phase unambiguously and converge most observables to high precision. To ensure that we only study differences in interacting physics, we study two representative states at $\nu=-1/2$ with (I) strong interactions $\epsilon_r =6$ and (II) weak interactions $\epsilon_r =60$ at $\theta=3.89^\circ$. Following the discussion in App. \ref{sec:app_band_projection}, we note that this point is adiabatically connected within the projected Hilbert space to the parameters $(\theta,\epsilon_r) = (3.7^\circ, 15)$ studied in the main text.

We begin by examining the occupations $n(\v{k}) = \braket{\hat{c}^\dagger_{\v{k}} \hat{c}_{\v{k}}}$ across the Brilouin zone. These are shown in Fig. \ref{fig:CFL_numerics}(a) of the main text. For $\epsilon_r=6$, the occupations are almost uniform, while for $\epsilon_r=60$, the occupations are highly non-uniform with a sharp jump emerging as bond dimension increases. This behavior is characteristic of a Fermi liquid in cylinder DMRG: since the system is quasi-1D, no true Fermi liquid with quasiparticle weight $Z>0$ can emerge. Instead, the ground state here is converging to (coupled) Luttinger liquids with a non-analyticity at the Fermi surface~\cite{varjas2013chiral}. The locations of these features for the weakly-interacting state form an almost-circular Fermi surface centered at $\Gamma$ --- precisely what is expected for the weakly interacting Fermi liquid, given that the top band has a minimum at $\Gamma$. So far we have established that the weakly interacting state is highly consistent with a Fermi liquid, but the uniform occupations of the strongly-interacting state are consistent with both trivial insulators or more exotic possibilities.

\begin{figure*}
   \centering
   \includegraphics[width=\linewidth]{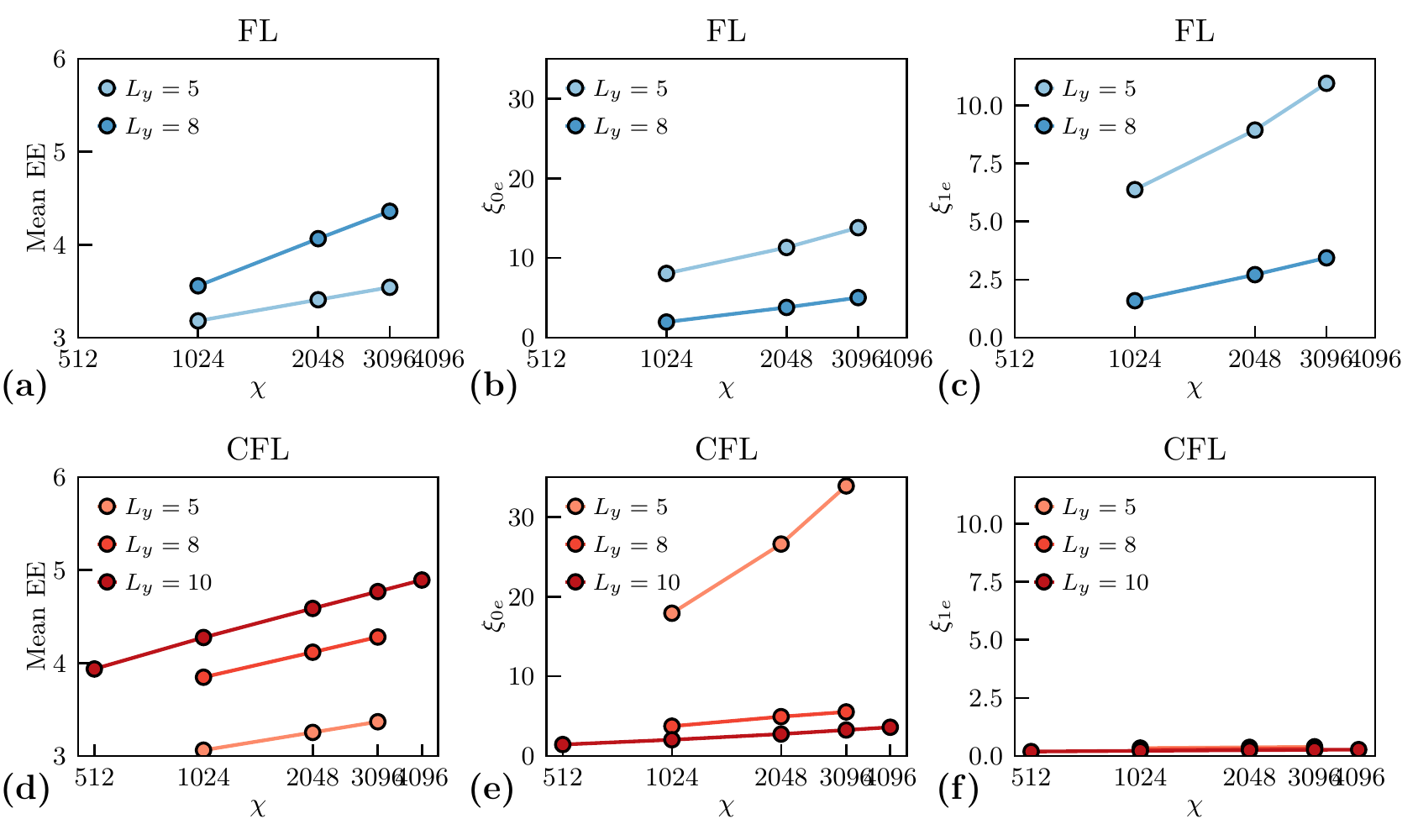}
   \caption{Entanglement properties as a function of bond dimension $\chi$. (a,d): the mean entanglement entropy measured. (b,e): the correlation length in the $Q_E=0$ sector. The divergence in correlation length is consistent with the presences of gapless neutral excitations, corresponding to the particle-hole excitations in the Fermi liquid and density fluctuations in the composite Fermi liquid. (c,f): the correlation length in the $Q_E=1$ sector. The Fermi liquid has a diverging correlation length, reflecting its gapless charged excitations -- the electrons. In contrast, the composite Fermi liquid has small correlations in this sector. Parameters: $(\theta,\epsilon_r,V) = (3.89^\circ,6,0)$.}    \label{fig:app_entanglement_correlation_length_scaling}
\end{figure*}

We now examine entanglement scaling and correlation lengths. Fig.~\ref{fig:app_entanglement_correlation_length_scaling}(a,d) shows the scaling of entanglement entropy as a function of bond dimension. We see that both the weakly and strongly interacting states are highly entangled, consistent with a gapless state. To determine the character of potential gapless modes, we examine the spectrum of the transfer matrix. For a Fermi liquid, where adding an electron, hole, or particle-hole pair are all gapless excitations, one expects electron-electron, hole-hole, and density-density correlations to all show power-law behavior, with a diverging correlation length. For $\epsilon_r =60$, we indeed find diverging correlation lengths for the electric charge $Q_E = +1,-1, 0$ sectors for all $k_y$, shown in Fig.~\ref{fig:app_entanglement_correlation_length_scaling}(b,c). We conclude that the weakly interacting state is indeed a Fermi liquid. For the strongly interacting state, we see a similar increase in $Q_E = 0$ correlations without any accompanying increase in $Q_E = \pm 1$ correlations, shown in Fig.~\ref{fig:app_entanglement_correlation_length_scaling}(e,f). This is consistent with a state that has gapless charge neutral excitations, but no gapless electrons and holes. 

\begin{figure*}
   \centering
   \includegraphics{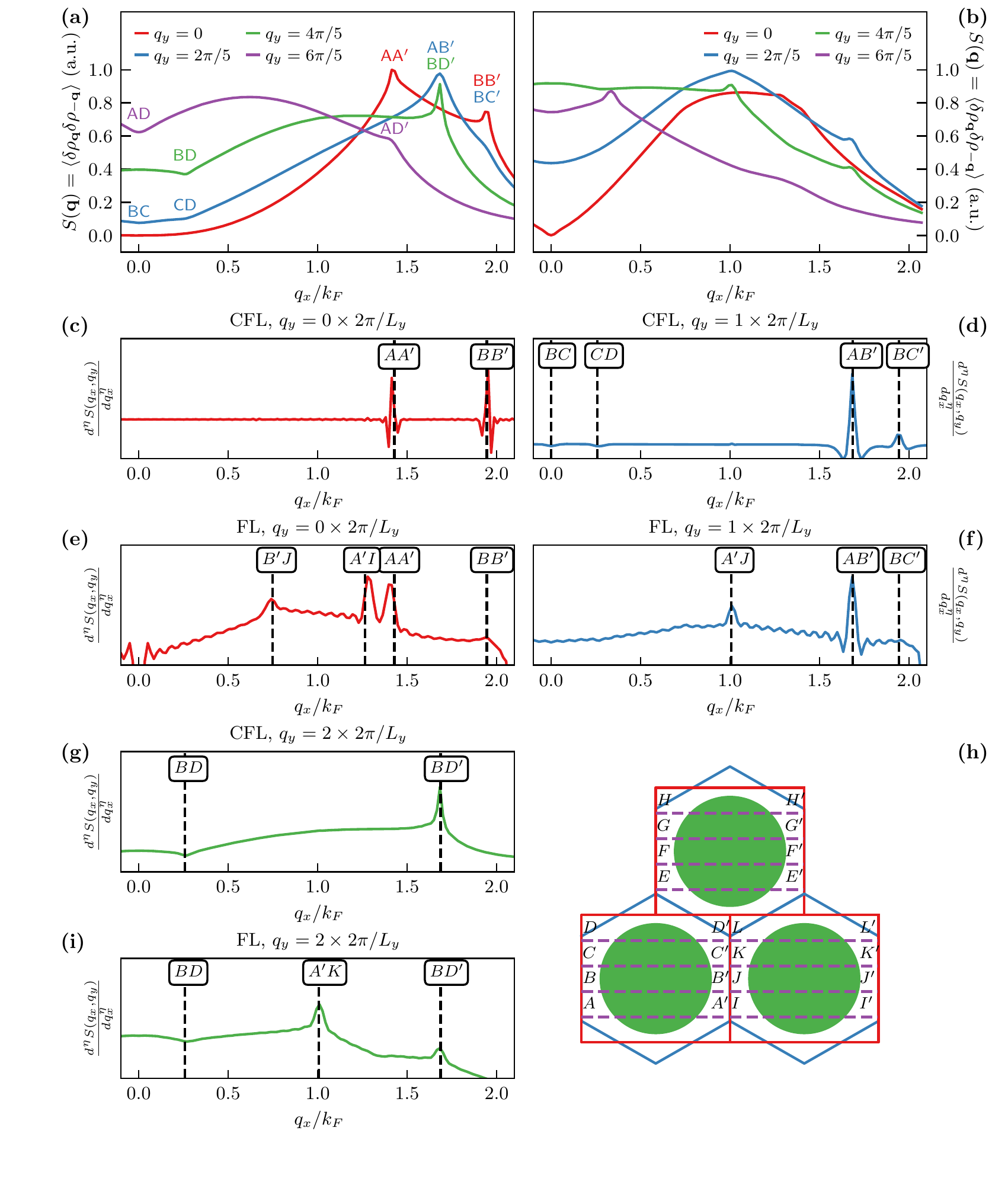}
   \caption{Quantitative comparison between FL and CFL structure factors with fermi surface scattering momenta. (a) Connected structure factor $S(\v{q})$ at constant $q_y$ as a function of $q_x$ for the CFL phase (also shown in main text). (b) Same for the FL at $\epsilon_r=60$. Fractional derivatives of $S(\v{q})$ with respect to $q_x$ to bring out detailed features are shown for $q_y =0$ (c,e), $q_y=1$ (d,f), and $q_y=2$ (g,i). Note that some peaks are only clearly visible for FL or CFL. In particular, \textit{umklapp} peaks between the first and higher Brillouin zones only clearly appear for the FL state. (h) Brillouin zone geometry and labelling scheme for vertical dashed lines. For instance, a line denoted `BD' corresponds to the $q_x$ position of $\v{q} = \v{k}_D - \v{k}_B$, where $\v{k}_B$ and $\v{k}_D$ are intersections between the wires and the Fermi surface. Parameters: $(\theta,\epsilon_r,V) = (3.89^\circ,6 \text{ or } 60,0)$.}
   \label{fig:app_structure_factor_details}
\end{figure*}

To understand the character of the neutral degrees of freedom, we turn to the structure factor $S(\v{q})$ described above. Fig.~\ref{fig:app_structure_factor_details} details the structure factor for both the weakly- and strongly-interacting states. As the minimum resolution in $q_x$ is set by the density-density correlation length, we focus on $L_y =5$ so that we may identify peaks, inflection points, and other sharp features.

We now show that the structure factors $S(\v{q})$ for both the weakly and strongly interacting states are consistent with an almost-circular Fermi surface. Explicitly, assume a Fermi surface (not necessarily of electrons) of half the Brillouin zone area, i.e. $k_F = \sqrt{A_{BZ}/2\pi}$. Examining the geometry, we see that four of the five wires intersect a putative Fermi surface. (The assumption that the momenta of the quasiparticles that make up the Fermi surface matches the momenta of the wires holds here, but not in general~\cite{GeraedtsScience}.) On this quasi-1D system, we expect features in $S(\v{q})$ whenever particle-hole pairs
\[
\hat{c}^\dagger_{\v{k}} \hat{c}_{\v{k}+\v{q}} \subset \hat{\rho}_{\v{q}}
\]
have $\v{k}$ and $\v{k}+\v{q}$ at intersections between wires and the Fermi surface. We label these intersections
\[
\v{k}_{A,B,C,D} = \left(-\sqrt{k_F^2 - k_y[n]^2}, k_y[n]\right), n = 1,2,3,4,
\]
on the left-hand side of the Fermi surface, and give corresponding labels $\v{k}_{A',B',\dots}$ to the right-hand Fermi surface and higher Brillouin zones (see Fig.~\ref{fig:app_structure_factor_details}). 

We identify features in $S(\v{q})$ by taking its fractional derivative following \cite{varjas2013chiral,GeraedtsScience}, defined through the Fourier transformations as follows:
\begin{equation}
   \frac{d^\eta}{dq_x^\eta}S(q_x,q_y) = \int dx e^{-iq_x x}|x|^\eta\left(\int dq_x' e^{iq_x' x}S(q_x',q_y)\right)
\end{equation}
The fractional derivatives magnifies inflections and discontinuities in higher derivatives, making the scattering processes much more visible to the eye. We take various $\eta \in [1/3,5/2]$ for display.

We observe that the FL structure factors show a much larger \textit{umklapp} scattering than the structure factors of CFL in Fig.~\ref{fig:app_structure_factor_details}. The reason is that the Fermi liquid has a large density modulation in the unit cell, strongly enhancing the amplitude of \textit{umklapp} scatterings, whereas the CFL has a much milder density modulation. We comment that enhancing the density modulation of the CFL may enhance the \textit{umklapp} scattering and introduce deformations of the composite Fermi surface~\cite{song2023CFL}.

\section{More on experimental aspect of Chern band CFL}

The CFL is a gapless, compressible phase and in the absence of a quantized response is more challenging to establish as compared to gapped FQHE states. Nevertheless, it has many striking physical responses which are distinct from that of other gapless phases, in particular the Fermi liquid (FL). The experimental signatures include transport properties, tunneling density of states and finite momentum conductivity.

A direct probe of the composite nature of the CFL quasiparticles is revealed in the tunneling density of states (tDOS). The tDOS for a Fermi liquid is finite even at zero bias, a consequence of the low energy electron like quasiparticles. However, physical electrons cannot be injected into a CFL at zero bias, leading to a ``pseudogap'' density of states. In particular the low energy tDOS  of a CFL varies as  $A(\omega)\propto e^{-\omega_0/\omega}$ as a function of bias $\omega$~\cite{STMCFLSong}. The tDOS can be measured through the tunneling current in two setups. Using scanning tunneling microscope (STM), electrons can be directly injected into the sample. Alternatively, the tunneling current between two CFLs separated by a tunneling barrier can be measured. The contrast in zero bias features have been shown directly in Ref.~\cite{STMCFLEisenstein}, in which the second setup is employed in the Landau level setting.

Transport measurements give distinct signatures for the CFL. If we denote $\tilde{\rho}$ the (dimensionless) resistivity   tensor for the CFL itself, in the half filled Landau level setting, the physical conductivity is given by: 
\begin{equation}
  \sigma = \frac{e^2}{2h} \left ({\bf \epsilon}  +\frac{\tilde{\rho}}{2} \right ),
  \label{Eqn:CFLsigmaSupp}
\end{equation}
where $\epsilon$ is the antisymmetric $2\times 2$ tensor. Now, in the absence of disorder and of scattering of the composite fermions, $\tilde{\rho}\rightarrow 0$, whch implies the  longitudinal conductivity $\sigma_{xx}$ vanishes. In contrast, in a clean Fermi liquid in a Chern band, the longitudinal conductivity is expected to {\it grow} as disorder is reduced~\cite{RevModPhys.82.1539}. One may further seek to measure thermoelectric transport signatures, encapsulated in the tensor ${\bf E} = \hat{S} \bf \nabla T $, where for the CFL in the half filled LL, $\hat{S} = \sigma^{-1} \epsilon \tilde{\rho}\tilde{\alpha}/2 $, where $\tilde{\alpha}$ is the thermoelectric conductivity of the  composite Fermi surface  Ref.~\cite{CFLTransport,WangCooperHalperinStern}. Finally, thermal conductivity measurements, although challenging, will serve as a smoking gun signature of the nFL, since the CFL will strongly violate the Wiedemann-Franz law~\cite{WangSenthilCFL}. 

Away from the CFL filling,  the composite fermions experience a  effective magnetic field tied to the filling deviation:  $\Delta \nu = \nu-1/2$, leading to Landau levels of the  composite Fermi surface. FCIs correspond to fully occupying these Landau levels\cite{JainCF89, HalperinLeeRead}, corresponding to fillings $\nu = -(p+1)/(2p+1)$,  and the many-body gap of the FCI is identified with the cyclotron gap of the Landau levels. The observation of FCIs at filling $\nu= -2/3, \,-3/5$ in moir\'e TMDs~\cite{FCITMD23_xiaodong, FCITMD23_Kinfai}, corresponding to $p=1, \,2$ is consistent with this expectation. On the other hand, an FL would not exhibit such quantum oscillations when the filling is changed. Analogously, imposing a one-dimensional periodic grating is expected to lead to geometric resonances on varying the filling $\Delta \nu$ in the moir\'e TMD setting, as have been observed for CFLs in Landau levels~\cite{HalperinLeeRead, WillettPRL,KangPRL}. 

A notable prediction of the CFL picture~\cite{HalperinLeeRead}, verified in experiments~\cite{SAWWillett} in the Landau level context, relates to the finite wave-vector conductivity  $\sigma_{xx}(\v{q})$ which can be probed by surface acoustic waves (SAWs). The CFL conductivity  is predicted to be {\em linear} in the wave-vector $|{\bf q} |$. This follows directly from the form of the CFL conductivity tensor \ref{Eqn:CFLsigmaSupp} combined with the fact that the intrinsic resistivity of the CFL in the clean limit is set by the finite wave-vector of the SAW probe $\tilde{\rho} \propto |{\v{q}}| $. 

Implementing the proposals above will require overcoming significant experimental challenges - from performing transport measurements on TMD moir\'e materials to working with the smaller size of moir\'e samples compared to GaAs/GaAlAs samples. Nevertheless, given the  experimental opportunity to study a zero field non-Fermi liquid which might give rise to of a non-Abelian topological phase, we anticipate there will be strong push to tackle these challenges.

\end{document}